\documentclass[prl,twocolumn,showpacs,amsmath,amssymb,superscriptaddress]{revtex4-2}
\usepackage{amsmath,amssymb,bm}
\usepackage{booktabs}
\newcommand{\T}{\mathcal{T}}
\usepackage{amsmath,amssymb,amsfonts,bm,color,graphicx}
\usepackage[unicode=true,colorlinks=true]{hyperref}
\usepackage[etex=true,export]{adjustbox}
\usepackage{makecell}
\usepackage{multirow}
\usepackage{verbatim}
\usepackage{longtable}
\hypersetup{linkcolor=blue, citecolor=blue,urlcolor=blue}
\usepackage{tabularx,array,diagbox}


\usepackage{booktabs}
\usepackage{amsbsy}

\usepackage{array}
\newcolumntype{L}[1]{>{\raggedright\arraybackslash}p{#1}}

\renewcommand{\arraystretch}{1.08}
 
\begin{document}
 
\title{\textcolor{red}{In-plane Anomalous Hall Effect in Unconventional Magnets}}
 
\author{Xun-Jiang Luo}
\email{xjluo@hmfl.ac.cn}
\altaffiliation{These authors contributed equally to this work.}
\affiliation{High Magnetic Field Laboratory, HFIPS, Chinese Academy of Sciences, Hefei, Anhui 230031, China}

\author{Yong-Ting Shi}
\altaffiliation{These authors contributed equally to this work.}
\affiliation{Institute of Applied Physics and Computational Mathematics, Beijing 100088, China}
\affiliation{High Magnetic Field Laboratory, HFIPS, Chinese Academy of Sciences, Hefei, Anhui 230031, China}

\author{Ding-Fu Shao}
\affiliation{Key Laboratory of Materials Physics, Institute of Solid State Physics,
HFIPS, Chinese Academy of Sciences, Hefei 230031, China}

\author{Ping Zhang}
\email{zhang\_ping@iapcm.ac.cn}
\affiliation{Institute of Applied Physics and Computational Mathematics, Beijing 100088, China}

\author{Ning Hao}
\email{haon@hmfl.ac.cn}
\affiliation{High Magnetic Field Laboratory, HFIPS, Chinese Academy of Sciences, Hefei, Anhui 230031, China}

\author{Mingliang Tian}
\email{tianml@hmfl.ac.cn}
\affiliation{High Magnetic Field Laboratory, HFIPS, Chinese Academy of Sciences, Hefei, Anhui 230031, China}

\begin{abstract}
\textcolor{red}{Unconventional magnets combine symmetry-compensated magnetic order with nonrelativistic spin splitting, calling for transport signatures that probe their magnetic phases. Here, we establish the in-plane anomalous Hall effect (IPAHE) as such a probe. We develop a unified framework connecting spin-space-group magnetic geometry to magnetic-point-group Hall response, treating spontaneous and field-induced IPAHE on the same footing. Applying it to all 122 magnetic point groups, we identify five groups supporting spontaneous IPAHE and 54 supporting field-induced IPAHE. Crucially, it reveals field-induced IPAHE in a large family of unconventional magnets, including altermagnets and odd-parity magnets. Under an in-plane rotating magnetic field, the Hall conductance exhibits one-, two-, or three-fold angular harmonics, providing a symmetry-resolved transport probe of unconventional magnetic order. In particular, within our classification, a twofold IPAHE pattern  necessarily implies $C_{4z}T$ symmetry, providing a direct transport probe of  tetragonal $d$-wave altermagnetism. A MAGNDATA screening identifies numerous candidates, including 55 altermagnetic or odd-parity-magnet candidates. First-principles calculations for the $d$-wave altermagnet CsCr$_2$S$_2$O and the $p$-wave magnet CeNiAsO explicitly confirm field-induced IPAHE. Our results establish a direct connection between IPAHE and unconventional magnetism, offering concrete material platforms and measurable transport signatures.}
\end{abstract}

\maketitle

\textit{Introduction}---Hall effects are fundamental to condensed matter physics \cite{Nagaosa2010,Sinova2015,RevModPhys.95.011002}. The Hall response is described by the Hall vector $\bm{\sigma}^{\mathrm{H}}$, which relates the Hall current $\bm{J}^{\mathrm{H}}$ to the electric field $\bm{E}$ through $\bm{J}^{\mathrm{H}}=\bm{E}\times\bm{\sigma}^{\mathrm{H}}$. In the conventional anomalous Hall (AH) effect, $\bm{\sigma}^{\mathrm{H}}$ is collinear with the magnetization $\bm{M}$, so that the AH current is orthogonal to both $\bm{E}$ and $\bm{M}$ \cite{Nagaosa2010}. \textcolor{red}{The in-plane anomalous Hall effect (IPAHE) departs from this paradigm in two distinct forms. Spontaneous IPAHE occurs when $\bm{M}$ lies in the current plane but $\bm{\sigma}^{\mathrm{H}}$ acquires a component perpendicular to $\bm{M}$ \cite{PhysRevB.72.104430,PhysRevB.109.155153,chen2025strain,Chen2025,kao2026lowdim,peng2026octupole,zheng2026cr3te4,peng2026cr12te2}. Field-induced IPAHE arises when an in-plane magnetic field generates an AH response orthogonal to the field, as proposed theoretically \cite{PhysRevB.84.085123,PhysRevLett.111.086802,qiaozhenghua2016,Liu2018,PhysRevB.100.165117,PhysRevB.103.214438,Battilomo2021,Sun2022,Cao2023,Miao2024Cd3As2,liu2025multipolar} and demonstrated experimentally in nonmagnetic, antiferromagnetic, and ferromagnetic materials \cite{Zhou2022,Nakamura2024,Lee2025EuZn2Sb2,Wang2025Co3Sn2S2,nishihaya2025sro,sankar2025fe3sn,Dai2026}. These advances raise a broader question: which magnetic phases can host IPAHE, and what physical information can its transport signatures reveal?}

\textcolor{red}{Unconventional magnetism provides a compelling setting for this question. Unconventional magnets combine symmetry-compensated magnetic order with nonrelativistic spin splitting, placing them beyond conventional ferromagnets and antiferromagnets \cite{ifmmode2022,BaiLing2024,Liu2025a,2025arXiv251005512L,2026arXiv260307643L,2026arXiv260521336L}. This rapidly expanding family includes altermagnets with even-parity spin splitting, odd-parity magnets with odd-parity spin splitting, and hybrid-parity magnets in which both coexist \cite{2026arXiv260307643L,2026arXiv260521336L}. Spin--orbital magnets provide another example, in which spin--orbit coupling (SOC) enables Hall responses forbidden by the spin-group symmetry \cite{Liu2026a}. Unconventional magnets have attracted widespread attention across condensed matter physics and materials science, while their experimental transport signatures remain limited. At the same time, IPAHE has been investigated on various material platforms, its occurrence in unconventional magnets remains largely unexplored. We therefore ask whether unconventional magnets can host IPAHE and, if so, whether its magnetic-field-angle dependence can provide transport signatures of the underlying magnetic phases.}

\begin{figure}[t]
\centering
\includegraphics[width=3.5in]{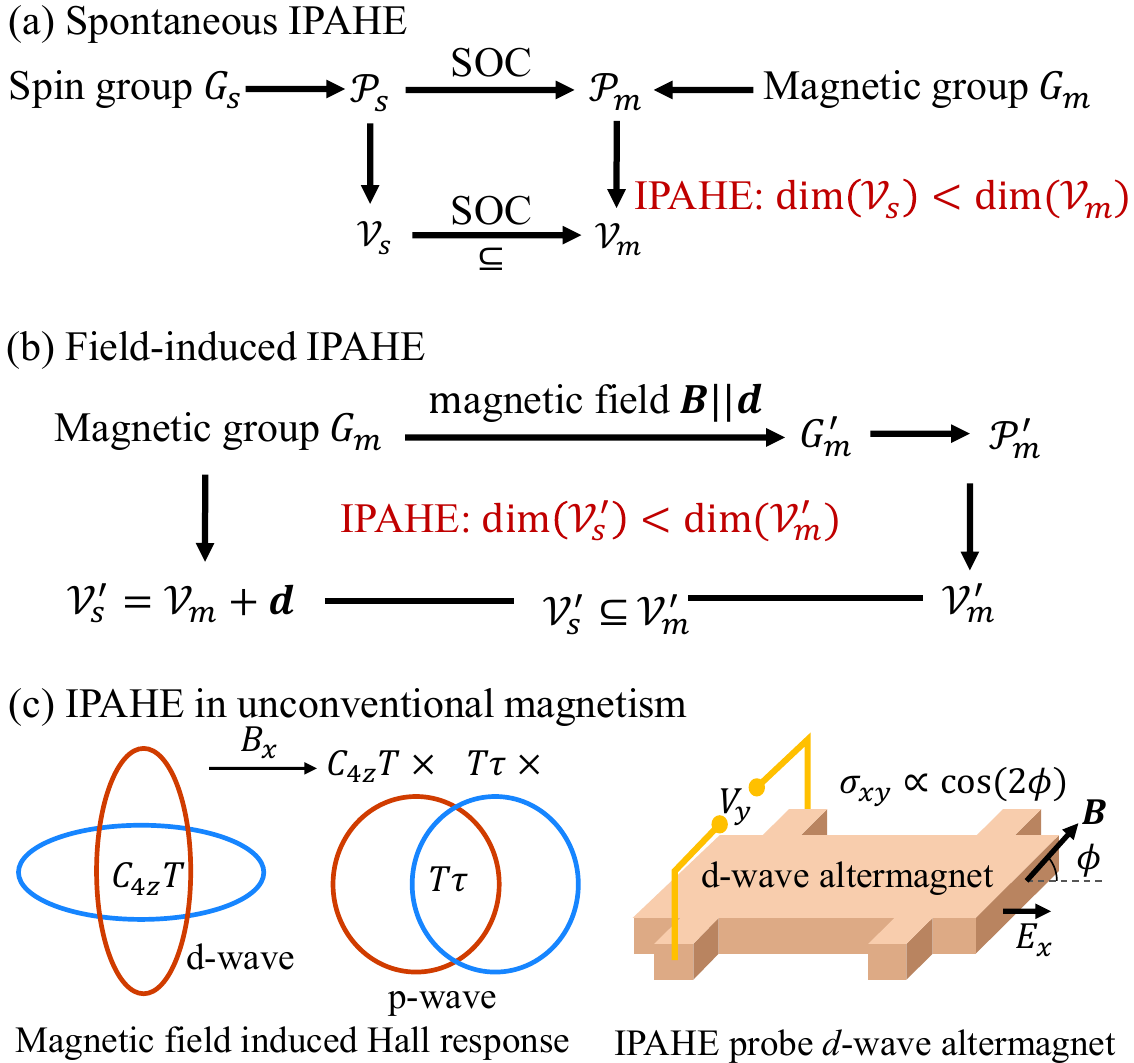}
\caption{\textcolor{red}{Unified symmetry framework for IPAHE and its realization in unconventional magnets. (a) Spontaneous IPAHE: the SSG $G_s$ and magnetic group $G_m$ determine the polarization spaces $\mathcal{V}_s$ and $\mathcal{V}_m$ through $\mathcal{P}_s$ and $\mathcal{P}_m$, respectively, with SOC connecting the two descriptions. IPAHE is allowed when $\dim(\mathcal{V}_s)<\dim(\mathcal{V}_m)$. (b) Field-induced IPAHE: a magnetic field along $\bm{d}$ lowers $G_m$ to $G_m'$, yielding $\mathcal{V}_s'=\mathcal{V}_m+\bm{d}$ and $\mathcal{V}_m'$. IPAHE requires $\dim(\mathcal{V}_s')<\dim(\mathcal{V}_m')$. (c) An in-plane field breaks the compensating $C_{4z}T$ and $T\tau$ symmetries, enabling IPAHE. IPAHE can be used to probe tetragonal $d$-wave altermagnetism.}}
\label{Fig1}
\end{figure}

\textcolor{red}{In this Letter, we establish IPAHE as an experimentally accessible probe of unconventional magnetism and develop a unified symmetry framework for its occurrence and angular response [Fig.~\ref{Fig1}].} The key insight is that magnetic geometry and AH response are governed by distinct but related symmetry structures. Magnetic geometry is described by the spin space group (SSG), where spin and spatial operations are decoupled \cite{LiuPengfei2022,CheXiaobing2024,JiangYi2024,XiaoZhenyu2024,Liu2026a}. Its spin operations form a point group $\mathcal{P}_s$, whose polarization space $\mathcal{V}_s$ determines the allowed net-magnetization directions. With SOC, the AH response is governed by the magnetic point group (MPG), whose spin operations form $\mathcal{P}_m$ with polarization space $\mathcal{V}_m$ determining the allowed directions of $\bm{\sigma}^{\mathrm{H}}$. We prove $\mathcal{V}_s\subseteq\mathcal{V}_m$, ensuring that a ferromagnetic moment along a polar direction permits a collinear AH response. Spontaneous IPAHE emerges when SOC-induced symmetry lowering yields $\dim(\mathcal{V}_s)<\dim(\mathcal{V}_m)$, allowing $\bm{\sigma}^{\mathrm{H}}$ to acquire components perpendicular to the magnetization [Fig.~\ref{Fig1}(a)].

\textcolor{red}{For field-induced IPAHE, an in-plane magnetic field along $\bm{d}$ lowers the MPG $G_m$ to $G_m'$, defining $\mathcal{V}_s'=\mathcal{V}_m+\bm{d}$ and the field-lowered Hall-response space $\mathcal{V}_m'$. IPAHE is allowed when $\dim(\mathcal{V}_s')<\dim(\mathcal{V}_m')$ [Fig.~\ref{Fig1}(b)]. Among all 122 MPGs, 54 satisfy this condition, including 40 nonferromagnetic groups [Table~\ref{MPGs}]. These include symmetries realized by altermagnets, odd-parity magnets, hybrid-parity magnets, and spin--orbital magnets, establishing IPAHE as a broad transport phenomenon in unconventional magnets. Under an in-plane rotating field, the Hall conductance exhibits one-, two-, or three-fold harmonics of the field angle $\phi$ [Table~\ref{MPGss}]. The MPG determines the leading harmonic and the symmetry-constrained extrema and nodes, yielding experimentally testable transport signatures of the underlying magnetic order. Importantly, within our classification, a twofold IPAHE pattern necessarily implies $C_{4z}T$ symmetry, providing direct transport prob for tetragonal $d$-wave altermagnetism [Fig.~\ref{Fig1}(c)]}

\textcolor{red}{Guided by these results, we screen the complete MAGNDATA database and identify candidates spanning ferromagnets, antiferromagnets, and unconventional magnets, including 55 altermagnetic or odd-parity-magnet candidates for field-induced IPAHE. First-principles calculations for the $d$-wave altermagnet CsCr$_2$S$_2$O \cite{10.1021/jacs.6c17601} and the $p$-wave magnet CeNiAsO \cite{BirkHellenes2023} provide explicit realizations in two distinct unconventional-magnet classes. By linking unconventional magnets to measurable Hall transport, our work connects the fields of IPAHE and unconventional magnetism and guides future experiments.}


\textit{Spontaneous IPAHE}---The magnetic geometry of a material is fundamentally described by its SSG, as detailed in the Supplemental Materials \cite{supp}. A general SSG operation is denoted by $g_s = \{U || R\}$, where $R$ and $U$ act separately on the spatial and spin degrees of freedom, respectively. The set of all spin operations $\{U\}$ in an SSG forms a point group $\mathcal{P}_s$, a finite subgroup of $O(3)$. A net magnetization $\bm{M}$ is allowed if and only if $\mathcal{P}_s$ is polar \cite{Liu2026a}, and its allowed polarization direction space can be expressed as
\begin{equation}
\mathcal{V}_s = \bigcap_{U \in \mathcal{P}_s} \ker(U - \mathbb{I}),
\end{equation}
where $\mathbb{I}$ is the $3\times 3$ identity matrix, and $\ker(U - \mathbb{I})$ denotes the null space of the matrix $(U - \mathbb{I})$, representing the spin directions invariant under the operation $U$. For a ferromagnet, this invariant subspace is nontrivial, satisfying $1 \le \dim(\mathcal{V}_s) \le 3$. Physically, the direction of the net magnetization $\bm{M}$ is strictly confined to $\mathcal{V}_s$.

When SOC is activated, the spin and spatial degrees of freedom are coupled, and the AH response is governed by the corresponding MPG. A general MPG operation takes the form $g_{m} = (\mathcal{R}, \eta)$, where $\mathcal{R}$ is a real-space point-group operation, and $\eta = +1$ ($-1$) corresponds to a unitary (antiunitary, i.e., combined with time reversal $\mathcal{T}$) operation. The Hall vector $\bm{\sigma}^{\mathrm{H}}$ is an axial vector and transforms under $g_m$ as
$A\bm{\sigma}^{\mathrm{H}}$ with $A=\eta\det(\mathcal{R})\mathcal{R}$. The set of matrices $\{A\}$ derived from an MPG forms a point group $\mathcal{P}_m$, representing the spin operations of the MPG. Analogous to the magnetization case, the symmetry-allowed directions of $\bm{\sigma}^{\text{H}}$ constitute the polarization space of $\mathcal{P}_m$,
\begin{equation}
\mathcal{V}_m = \bigcap_{A \in \mathcal{P}_m} \ker(A - \mathbb{I}).
\end{equation}
Physically, the direction of $\bm{\sigma}^{\text{H}}$ is strictly confined to $\mathcal{V}_m$.

Because the magnetic space group is a subgroup of the SSG \cite{Liu2026a}, its spin-space representation satisfies $\mathcal{P}_m \subseteq \mathcal{P}_s$. Consequently, any vector $\bm{v} \in \mathcal{V}_s$ remains invariant under all operations in $\mathcal{P}_m$, yielding a fundamental subspace relation:
\begin{equation}
\mathcal{V}_s \subseteq \mathcal{V}_m.
\end{equation}
Physically, this guarantees that a ferromagnetic moment along $\bm{v} \in \mathcal{V}_s$ necessarily permits a collinear AH response, namely $\bm {\sigma}^{\text{H}}\cdot \bm{v}\neq 0$. Since SOC lowers the symmetry from the spin group to the magnetic group, it relaxes the symmetry constraints, potentially enlarging the AH response space $\mathcal{V}_m$ relative to the magnetization space $\mathcal{V}_s$. Therefore, spontaneous IPAHE can emerge from this symmetry mismatch when
\begin{equation}
\mathcal{V}_s \subsetneq \mathcal{V}_m \quad \left[\dim(\mathcal{V}_m) > \dim(\mathcal{V}_s) > 0\right].
\label{eq4}
\end{equation}
This condition allows $\bm {\sigma}^{\text{H}}$ to acquire components perpendicular to the net magnetization, namely $\bm {\sigma}^{\text{H}}\times \bm{M}\neq 0$.

\begin{table*}[t]
\centering
\setlength\tabcolsep{2pt}
\renewcommand{\arraystretch}{1.45}
\caption{Spontaneous and field-induced IPAHE in magnetic materials. Among the 122 MPGs, 5 and 54 can support spontaneous and field-induced IPAHE, respectively. They are classified into ferromagnetic and non-ferromagnetic point group cases, describing various magnetic phases. The last column presents the number of candidate materials identified from the MAGNDATA database. }
\label{MPGs}
\begin{tabular}{|c|c|c|c|c|}
\hline
IPAHE & MPG class & MPGs & Phases & Materials \\
\hline
Spontaneous & Ferromagnetic 
& $2'$, $m'$, $2'/m'$, $1$, $\bar{1}$ 
& Ferromagnets & 33 \\
\hline
\multirow{2}{*}{Field-induced} 
& Ferromagnetic 
& \makecell{
$2$, $m$, $2/m$, $4$, $\bar{4}$, $4/m$, $3$, $\bar{3}$, $6$, $\bar{6}$, $6/m$,\\
$32'$, $3m'$, $\bar{3}m'$
} 
& \makecell{Ferromagnets,\\ spin-orbital magnetism} 
& 131 \\
\cline{2-5}
& Non-ferromagnetic 
& \makecell{
$32$, $3m$, $\bar{3}m$, $1.1'$, $1.\bar{1}'$, $1'$, $2.1'$, $m.1'$, \\ $2/m.1'$, $4.1'$, $\bar{4}.1'$,
$4/m.1'$, $3.1'$, $\bar{3}.1'$, $32.1'$, $3m.1'$, \\ $6.1'$, $\bar{6}.1'$, $6/m.1'$, $2'/m$, $2/m'$,
$4'$, $\bar{4}'$, $4'/m$,\\ $4/m'$, $4'm'm$, $\bar{4}'2'm$, $4'/mm'm$, $3'$, $\bar{3}'$, $\bar{3}'m$,
$\bar{3}'m'$, \\ $6'/m$, $6/m'$, $6'/m'$, $6'22'$, $6'mm'$, $\bar{6}'m2'$, $6'/m'mm'$, $\bar{6}'m'2$
} 
& \makecell{ Antiferromagnets,\\ unconventional magnets } 
& 482 \\
\hline
\end{tabular}
\end{table*}

Applying this framework to ferromagnets, we find that the subspace relation guarantees $\dim(\mathcal{V}_m) \ge \dim(\mathcal{V}_s) \ge 1$. Therefore, spontaneous IPAHE requires $\dim(\mathcal{V}_m) \ge 2$ and $\dim(\mathcal{V}_m) > \dim(\mathcal{V}_s)$. Evaluating $\mathcal{V}_m$ for all 31 ferromagnetic MPGs reveals three cases:
\begin{itemize}
\item {1D $\mathcal{V}_m$ (26 groups):} The AH vector $\bm{\sigma}^{\text{H}}$ is strictly collinear with $\bm{M}$, yielding only the conventional AH effect.
\item {2D $\mathcal{V}_m$ (3 groups: $2'$, $m'$, $2'/m'$):} $\mathcal{V}_m$ spans a plane. If $\dim(\mathcal{V}_s)=1$ (e.g., in group $2'$, $\mathcal{V}_s$ is the $x$-axis while $\mathcal{V}_m$ is the $x$-$z$ plane), a magnetization along $\hat{x}$ permits an AH response spanning both $\hat{x}$ and $\hat{z}$, realizing IPAHE. If $\dim(\mathcal{V}_s)=2$, only the conventional AH effect occurs.
\item {3D $\mathcal{V}_m$ (2 groups: $1$, $\bar{1}$):} $\mathcal{V}_m = \mathbb{R}^3$. IPAHE can emerge if $\dim(\mathcal{V}_s) < 3$. If $\dim(\mathcal{V}_s)=3$, only the conventional AH effect occurs.
\end{itemize}
This classification provides a definitive design principle: materials in the $2'$, $m'$, $2'/m'$, $1$, or $\bar{1}$ point groups exhibit spontaneous IPAHE provided that $\dim(\mathcal{V}_s) < \dim(\mathcal{V}_m)$. Remarkably, our framework naturally unifies the recently proposed spin-orbital magnetism \cite{Liu2026a,supp}, in which the net magnetization vanishes ($\bm{M}=0$) under the SSG but remains finite under the MPG. In our framework, this corresponds to $\dim(\mathcal{V}_s) = 0$ with $\dim(\mathcal{V}_m) \ge 1$, thereby enabling AH effects in antiferromagnets. 

\textit{Field-Induced IPAHE}---We now consider IPAHE induced by an external field $\bm{B} \parallel \bm{d}$ in a system with an initial MPG $G_m$. As an axial vector, $\bm{B}$ breaks time-reversal symmetry and lowers spatial symmetry, preserving only operations that leave $\bm{B}$ invariant. The surviving operations form the group
\begin{equation}
\mathcal{S}_{\bm{d}} = \{I, C_{n\bm{d}}, m_{\bm{d}}, m'_{\bm{p}}, C'_{2\bm{p}}\},
\end{equation}
where $I$ is inversion, $C_{n\bm{d}}$ is the $n$-fold rotation about $\bm{d}$, and $m_{\bm{d}}$ is the mirror plane perpendicular to $\bm{d}$. $m'_{\bm{p}}$ and $C'_{2\bm{p}}$ denote the antiunitary operations that combine time reversal with a mirror plane perpendicular to $\bm{p}$ and the twofold rotation about $\bm{p}$ ($\bm{p} \perp \bm{d}$), respectively. The field-lowered MPG is the intersection of $G_m$ and $\mathcal{S}_{\bm{d}}$:
\begin{equation}
	G_m' = G_m \cap \mathcal{S}_{\bm{d}},
\end{equation}
which is the maximal common subgroup of $G_m$ and $\mathcal{S}_{\bm{d}}$.

This symmetry reduction establishes the condition for field-induced IPAHE. The constrained polarization space, combining the initially allowed AH response space $\mathcal{V}_m$ with the field direction $\bm{d}$, is $\mathcal{V}'_s = \mathcal{V}_m + \bm{d}$, with dimension $\dim(\mathcal{V}'_s) = \dim(\mathcal{V}_m) + \alpha$ ($\alpha = 1$ if $\bm{d} \notin \mathcal{V}_m$, and $0$ otherwise). The field-lowered MPG $G_m'$ defines the new AH response space $\mathcal{V}'_m$. Similarly, we have $\mathcal{V}_s^{'} \subseteq \mathcal{V}_m^{'}$ \cite{supp} and field-induced IPAHE requires a strict dimensional mismatch:
\begin{equation}
\dim(\mathcal{V}'_m) > \dim(\mathcal{V}'_s).
\label{eq7}
\end{equation}
When this inequality is satisfied, there exist elements in $\mathcal{V}'_m$ orthogonal to both the initially allowed Hall response directions and $\bm{d}$, thereby allowing a field-induced Hall response orthogonal to the applied field $\bm{B}$.

%

We apply our framework to all 122 MPGs---32 colorless, 32 gray, and 58 black-and-white---to identify those capable of hosting field-induced IPAHE. For each MPG $G_m$, we consider magnetic fields along the three orthogonal axes of the conventional Cartesian coordinate system, evaluating the surviving operations that form $G_m'$, as detailed in the SMs \cite{supp}. By comparing $G_m$ and $G_m'$, we obtain $\dim(\mathcal{V}'_m)$ and $\dim(\mathcal{V}'_s)$, allowing us to identify the MPGs that can support field-induced IPAHE \cite{supp}. Specifically, we find that 14 of the 31 ferromagnetic MPGs and 40 of the 91 non-ferromagnetic MPGs can support IPAHE. These 40 non-ferromagnetic MPGs include 16 gray groups describing nonmagnetic materials \cite{PhysRevB.103.214438}. In total, 54 MPGs can host field-induced IPAHE, as summarized in Table~\ref{MPGs}. In the SMs, we list the symmetry-allowed $\bm{\sigma}^{\text{H}}$ under $G_m$ and $G_m'$, respectively, for each direction of the magnetic field \cite{supp}.

Remarkably, while our preceding analysis of field-induced IPAHE is rooted in MPGs, integrating the SSG significantly enriches the magnetic phases that can support field-induced IPAHE. Within the SSG framework, unconventional magnets---which exhibit spin splitting while retaining an antiferromagnetic geometry---can emerge. These unconventional magnetic materials can host field-induced IPAHE as long as their corresponding MPGs belong to those listed in Table~\ref{MPGs}. 
This scenario is realized in two distinct cases. The first involves unconventional magnets described by the ferromagnetic point groups listed in Table~\ref{MPGs}, which fall into the category of spin-orbital magnetism \cite{Liu2026a}. The second involves unconventional magnets described by the non-ferromagnetic point groups listed in Table~\ref{MPGs}, which lack space-time-reversal ($PT$) symmetry. Consequently, our framework not only unifies previously studied field-induced IPAHE in nonmagnetic \cite{PhysRevLett.111.086802,PhysRevB.103.214438}, ferromagnetic \cite{PhysRevB.109.155153}, and antiferromagnetic materials \cite{Cao2023}, but also predicts a broader family of unconventional magnets---including altermagnets, odd-parity magnets, and hybrid-parity magnets---that realize IPAHE \cite{ifmmode2022,Liu2025a,2025arXiv251005512L,2026arXiv260307643L,2026arXiv260521336L}. The MPGs for these phases are summarized in Table~\ref{MPGs}.

\textit{Field dependence of IPAHE}---For the field-induced IPAHE, the Hall
conductance depends on the direction of the applied magnetic field. When the
field rotates within the $xy$ plane,
$\mathbf B(\phi)=B(\cos\phi\,\hat x+\sin\phi\,\hat y)$ with $\phi$ being the angle between $\mathbf B$ and the $x$ axis, the Hall
conductance $\sigma_{xy}$ can generally be expanded as
\begin{align}
\sigma_{xy}(\phi)
&=\sum_{n=1}^{\infty}\bigl[a_n\cos(n\phi)+b_n\sin(n\phi)\bigr].
\label{eq:fourier}
\end{align}
where we have fixed the magnitude of $B$. Remarkably, the leading harmonic of $\sigma_{xy}(\phi)$ is completely
determined by the MPG $G_m$. A symmetry operation $g_m$ that keeps
the $xy$ plane invariant maps the field angle $\phi$ to $\phi_g$ and imposes
the constraint \cite{supp}
\begin{align}
\sigma_{xy}(\phi_g)=\xi\,\sigma_{xy}(\phi),
\label{eq:rule1}
\end{align}
where $\xi=(-1)^{\eta}\det(\mathcal R)(\mathcal R)_{zz}=\pm1 $.


Applying Eq.~(\ref{eq:rule1}) to the 28 MPGs that can host a $z$-component Hall response
induced by rotating the magnetic field within the $xy$ plane, the leading
angular dependence fall into the six symmetry classes summarized in
Table~\ref{MPGss}. The one-fold, two-fold, and three-fold harmonics
originate from the monoclinic/triclinic, tetragonal, and trigonal symmetry
constraints, respectively. Such field-angle-dependent Hall
responses have been experimentally observed in EuCd$_2$Sb$_2$
\cite{Nakamura2024} and Fe$_3$Sn \cite{sankar2025fe3sn}. The complete
symmetry analysis and the corresponding results for the other rotation
planes are given in the SMs \cite{supp}.

{\color{red}Combining angular periodicity with symmetry-enforced node positions across different rotation planes can substantially narrow the compatible MPGs and reveal the underlying magnetic symmetry, thereby providing a transport probe of unconventional magnetism. Crucially, within our classification [Table.~\ref{MPGss}], the observation of a twofold IPAHE pattern necessarily implies the presence of $C_{4z}T$ symmetry. This provides a direct probe of $d$-wave altermagnetism, whose defining feature is that opposite-spin sublattices are related by a fourfold rotation, meaning that the system hosts the $C_{4z}T$ operation. Therefore, once the collinear magnetic order is independently established, the twofold IPAHE periodicity constitutes smoking-gun transport evidence for the $d$-wave altermagnetic phase.}

\begin{table}[t]
\centering
\setlength\tabcolsep{1pt}
\renewcommand{\arraystretch}{1.45}
\caption{Leading angular dependences of the field-induced Hall conductance
$\sigma_{xy}(\phi)$, classified by crystal system, for the 28 MPGs among the
54 IPAHE-supporting MPGs that can host a $z$-component Hall response under an
in-plane magnetic field.}
\label{MPGss}
\begin{tabular}{|c|c|c|}
\hline
{Crystal system} & {Magnetic point groups} & \textbf{$\sigma_{xy}(\phi)$} \\
\hline
Triclinic &
\makecell[l]{$1.1'$, $\bar{1}.1'$, $\bar{1}'$}
& $a_{1}\cos\phi+b_{1}\sin\phi$ \\
\hline
Monoclinic &
\makecell[l]{$2$, $2.1'$, $m$, $m.1'$, \\ $2/m$,
  $2/m.1'$, $2'/m$, $2/m'$}
& $a_{1}\cos\phi$ \\
\hline
\multirow{2}{*}{Tetragonal} &
\makecell[l]{$4'$, $\bar{4}'$, $4'/m$}
& $a_{2}\cos2\phi+b_{2}\sin2\phi$ \\
\cline{2-3}
 &
\makecell[l]{$4'm'm$, $\bar{4}'2'm$, $4'/mm'm$}
& $a_{2}\cos2\phi$ \\
\hline
\multirow{2}{*}{Trigonal} &
\makecell[l]{$3.1'$, $\bar{3}.1'$, $\bar{3}'$}
& $a_{3}\cos3\phi+b_{3}\sin3\phi$ \\
\cline{2-3}
 &
\makecell[l]{$32$, $32.1'$, $3m$,  $3m.1'$,\\ $\bar{3}m$,
 $\bar{3}m.1'$, $\bar{3}'m$, $\bar{3}'m'$}
& $b_{3}\sin3\phi$ \\
\hline
\end{tabular}
\end{table}

\begin{figure*}[t]
\centering
\includegraphics[width=7in]{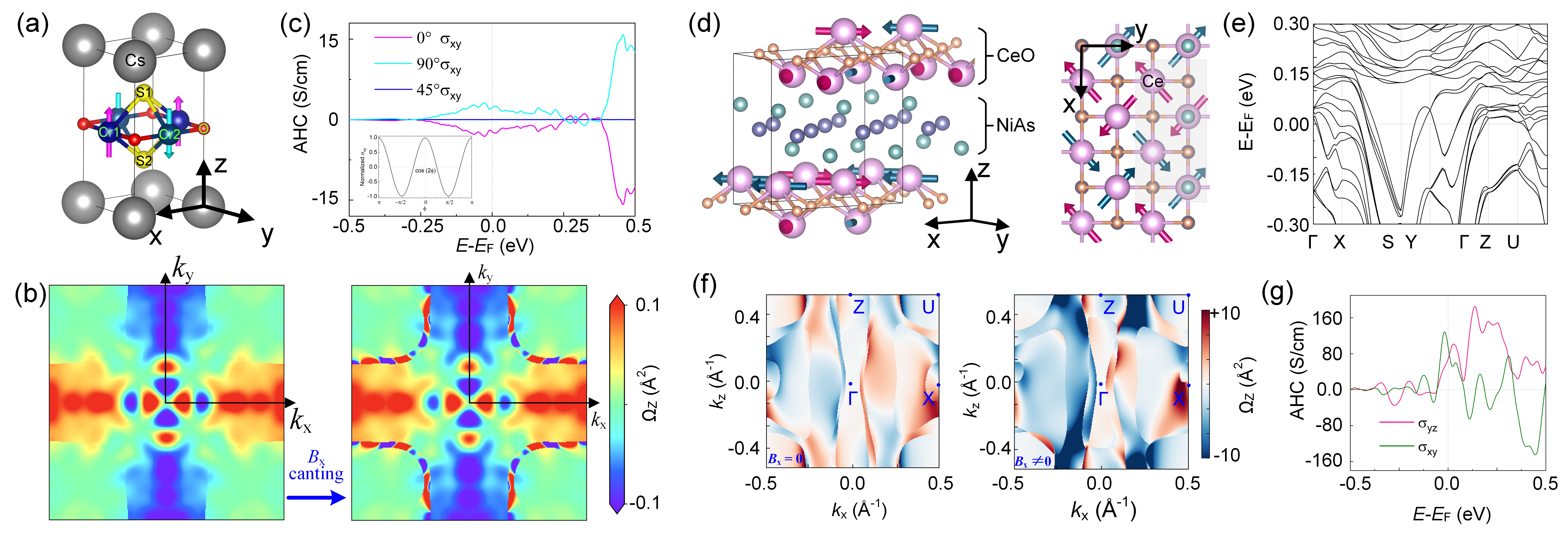}
\caption{\textcolor{red}{Field-induced IPAHE in CsCr$_2$S$_2$O and CeNiAsO.
(a) Magnetic structure of CsCr$_2$S$_2$O.
(b) Berry curvature distribution of $\Omega_z$ in the $k_z=0$ plane without (left) and with (right) spin canting.
(c) $\sigma_{xy}$ at $\phi=0^\circ$, $45^\circ$, and $90^\circ$. The sign reversal and the suppressed response at $45^\circ$ follow $\sigma_{xy}\propto\cos(2\phi)$ (inset).
(d) Magnetic structure and (e) band structure of CeNiAsO.
(f) Berry curvature $\Omega_z$ in the $k_y=0$ plane without (left) and with (right) spin canting.
(g) $\sigma_{xy}$ and $\sigma_{yz}$ of CeNiAsO with spin canting.
All calculations include SOC. The constrained in-plane moment is $0.1\,\mu_{\mathrm B}$ per magnetic atom.}}
\label{Fig2}
\end{figure*}

\textit{Candidate materials and first-principles calculations}---
Based on the established symmetry framework, we screen the MAGNDATA database \cite{GallegoMagndataI} and identify 33 candidate ferromagnets that can host spontaneous IPAHE, as detailed in the SMs \cite{supp}. For the field-induced case, 78 candidate ferromagnets are found to support IPAHE. Furthermore, 53 candidate spin-orbital magnets, categorized as unconventional magnets, are also predicted to exhibit field-induced IPAHE. In total, 131 materials described by ferromagnetic point groups can realize field-induced IPAHE. Turning to materials characterized by non-ferromagnetic point groups, we identify 482 candidate materials capable of hosting field-induced IPAHE, among which 55 are altermagnets or odd-parity magnets \cite{supp}. A summary of the candidate material counts is presented in Table~\ref{MPGs}.

\textcolor{red}{
To validate the established symmetry framework in unconventional magnets, we first consider the experimentally characterized $d$-wave altermagnet candidate CsCr$_2$S$_2$O [Fig.~\ref{Fig2}(a)] \cite{10.1021/jacs.6c17601}. Its high-temperature (326 K) tetragonal phase hosts an experimentally characterized C-type antiferromagnetic order with Cr moments along the $c$ axis, corresponding to the MPG $4'/mm'm$ \cite{10.1021/jacs.6c17601}. First-principles calculations reveal the expected $d$-wave spin splitting \cite{10.1021/jacs.6c17601}. With SOC included, the zero-canting Berry curvature $\Omega_z$ exhibits a quadrupolar distribution enforced by $C_{4z}T$ symmetry, and its Brillouin-zone integral vanishes [Fig.~\ref{Fig2}(b), left panel]. An in-plane field along $x$ lowers the MPG to $2'/m'$, allowing a Hall vector in the $x$--$z$ plane. Consistently, an $x$-directed constrained spin canting redistributes $\Omega_z$ [Fig.~\ref{Fig2}(b), right panel] and produces a finite $\sigma_{xy}$. As shown in Fig.~\ref{Fig2}(c), calculations at $\phi=0^\circ$, $45^\circ$, and $90^\circ$ reveal opposite signs of $\sigma_{xy}$ at $0^\circ$ and $90^\circ$ and a nearly vanishing response at $45^\circ$, consistent with the predicted $\sigma_{xy}(\phi)\propto\cos(2\phi)$ dependence.}

\textcolor{red}{We next demonstrate field-induced IPAHE in the $p$-wave magnet CeNiAsO [Fig.~\ref{Fig2}(d)], whose electronic band structure with SOC is shown in Fig.~\ref{Fig2}(e). Its zero-field MPG is $2.1'$, for which the AH response is forbidden. An $x$-directed in-plane field lowers the MPG to $2'$, with $C_{2y}T$ as the remaining point-group operation. This lowered symmetry permits $\sigma_{yz}$ and $\sigma_{xy}$ while requiring $\sigma_{zx}=0$. With SOC included, $\Omega_z$ in the $k_y=0$ plane is locally finite in the zero-canting state but integrates to zero over the Brillouin zone [Fig.~\ref{Fig2}(f), left panel]. A constrained $x$-directed canting of approximately $0.1\,\mu_{\mathrm B}$ redistributes the Berry curvature [Fig.~\ref{Fig2}(f), right panel] and produces nonzero $\sigma_{xy}$ [Fig.~\ref{Fig2}(g)]. In CeNiAsO, the emergence of IPAHE directly signals the field-induced breaking of the time-reversal--translation ($T\tau$) symmetry, which is important for realizing $p$-wave magnet \cite{BirkHellenes2023}. Moreover, the symmetry-allowed Berry-curvature dipole in the zero-canting state can generate a second-order nonlinear Hall response \cite{PhysRevB.107.115142}, providing a complementary transport probe of the magnetic order in CeNiAsO.}

\textit{Conclusion and discussion}---
\textcolor{red}{In summary, we have established a definition-level symmetry framework that unifies spontaneous and field-induced IPAHE by connecting the magnetic geometry described by the SSG to the Hall response governed by the MPG. The framework provides a complete classification of all 122 MPGs, predicts the angular dependence of field-induced IPAHE, and enables a database-wide search of magnetic materials. It reveals IPAHE in broad classes of unconventional magnets and establishes IPAHE as an experimentally accessible transport probe of these magnetic phases. First-principles calculations on the concrete materials provide examples of altermagnets and $p$-wave magnets.}

\textcolor{red}{Altermagnetism is defined at the symmetry level: opposite-spin sublattices are related by a crystallographic rotation or mirror operation, rather than by translation or inversion \cite{ifmmode2022}. Existing experimental observations, such as momentum-dependent spin splittings measured by ARPES \cite{Jiang2025DwaveKV2Se2O,Zhang2025SpinValleyRbV2Te2O,Hu2026SpinValleyNodalLines}, provide important indications of altermagnetism; however, similar splittings can also arise from SOC and therefore do not by themselves establish the defining altermagnetic symmetry. By contrast, within our classification, a twofold IPAHE pattern in a tetragonal magnet necessarily implies $C_{4z}T$ symmetry. Once collinear magnetic order is independently established, this periodicity constitutes smoking-gun transport evidence for the $d$-wave altermagnetic phase. This signature can be directly tested by angle-resolved IPAHE measurements in widely studied candidates, including RuO$_2$ \cite{Fedchenko2024RuO2},  KV$_2$Se$_2$O \cite{Jiang2025DwaveKV2Se2O}, and CsCr$_2$S$_2$O \cite{10.1021/jacs.6c17601}, providing a concrete route for the experimental identification of $d$-wave altermagnet. Notably, a third-order nonlinear Hall response has already been reported in RuO$_2$ \cite{rv1n-vr4p}; combining this observation with a future measurement of twofold IPAHE would single out the $4'/mm'm$ MPG \cite{PhysRevB.107.115142}  and provide a strong test of its proposed $d$-wave altermagnetic phase. }


We emphasize that the classification in Table~\ref{MPGs} is restricted to conventional measurement geometries, in which the magnetic field is applied along one of the three orthogonal Cartesian axes and the Hall response is characterized by $\sigma_{xy}$, $\sigma_{xz}$, and $\sigma_{yz}$. Within this scope, our analysis provides a complete classification of the MPGs supporting field-induced IPAHE. For a generic field direction $\boldsymbol{d}$, however, the surviving subgroup $G_m'(\boldsymbol{d})$ and the corresponding Hall-response space $\mathcal{V}_m'(\boldsymbol{d})$ may differ from those obtained for fields along the Cartesian axes. Consequently, if both the field orientation and the Hall-detection geometry are allowed to vary continuously, additional MPGs beyond those listed in Table~\ref{MPGs} can support field-induced IPAHE \cite{supp}.

\section{Acknowledgment}
Xun-Jiang Luo thanks helpful discussions with Jianhui Zhou and Xilin Feng. This work was supported by National Key R\&D Program of China (Grants No.
2022YFA1403200 and No.2024YFA1613200), National
Natural Science Foundation of China (Grants No.
92265104), the Basic Research Program
of the Chinese Academy of Sciences Based on Major
Scientific Infrastructures (Grant No. JZHKYPT-
2021-08), the CASHIPS Directors Fund (Grant
No. BJPY2023A09), Anhui Provincial Major S\&T
Project(s202305a12020005), and the High Magnetic
Field Laboratory of Anhui Province under Contract No.
AHHM-FX-2020-02.
 
\bibliography{reference,references1,ref}

@article{GallegoMagndataI,
	author = {Gallego, Samuel V. and Perez-Mato, J. Manuel and Elcoro, Luis and Tasci, Emre S. and Hanson, Robert M. and Aroyo, Mois I. and Madariaga, Gotzon},
	title = {{\it MAGNDATA}: towards a database of magnetic structures. II. The incommensurate case},
	journal = {J. Appl. Cryst.},
	year = {2016},
	volume = {49},
	number = {6},
	pages = {1941--1956},
	month = {12},
	doi = {10.1107/S1600576716015491},
	url = {https://doi.org/10.1107/S1600576716015491}
}

@article{10.1021/jacs.6c17601,
    author = {Liu, Yi and Xu, Chen-Chao and Bao, Jin-Ke and Lv, Bai-Jiang and Gao, Zi-Hang and Li, Hao and Li, Jing and Lin, Yi-Qiang and Li, Hua-Xun and Lu, Yi-Ming and Zhao, Xin-Yu and Yang, Wu-Zhang and Zhang, Zhen-Yi and Chen, Xian-Yan and Jiao, Wen-He and Liu, Ji-Yong and Wu, Si-Qi and Shen, Fei-Ran and He, Lun-Hua and Zhu, Bai-Ren and Jin, Wen-Tao and Cao, Guang-Han},
    title = {Altermagnetism
and Room-Temperature Metal-to-Insulator
Transition in CsCr2S2O},
    journal = {Journal of the American Chemical Society},
    volume = {148},
    number = {38},
    pages = {41707-41716},
    year = {2026},
    month = {09},
    issn = {0002-7863},
    doi = {10.1021/jacs.6c17601},
    url = {https://doi.org/10.1021/jacs.6c17601},
}

@article{PerezMatoBCS2015,
	author = {Perez-Mato, J.M. and Gallego, S.V. and Tasci, E.S. and Elcoro, L. and de la Flor, G. and Aroyo, M.I.},
	title = {Symmetry-Based Computational Tools for Magnetic Crystallography},
	journal = {Annual Review of Materials Research},
	volume = {45},
	pages = {217--248},
	number = {Volume 45, 2015},
	DOI = {https://doi.org/10.1146/annurev-matsci-070214-021008},
	year = {2015},
	type = {Journal Article}
}

@article{KresseFurthmuller1996,
  author = {Kresse, G. and Furthm\"uller, J.},
  title = {{E}fficient iterative schemes for ab initio total-energy calculations using a plane-wave basis set},
  journal = {Phys. Rev. B},
  volume = {54},
  pages = {11169--11186},
  number = {0},
  DOI = {10.1103/PhysRevB.54.11169},
  year = {1996},
  type = {Journal Article}
}

@article{KresseJoubert1996,
author = {G. Kresse and J. Furthm\"uller},
title = {{E}fficiency of ab-initio total energy calculations for metals and semiconductors using a plane-wave basis set},
journal = {Comput. Mater. Sci.},
volume = {6},
number = {1},
pages = {15--50},
ISSN = {0927--0256},
DOI = {https://doi.org/10.1016/0927-0256(96)00008-0},
year = {1996},
type = {Journal Article}
}

@article{KresseG1999,
  author = {Kresse, G. and Joubert, D.},
  title = {{F}rom ultrasoft pseudopotentials to the projector augmented-wave method},
  journal = {Phys. Rev. B},
  volume = {59},
  pages = {1758--1775},
  number = {0},
  DOI = {10.1103/PhysRevB.59.1758},
  url = {https://link.aps.org/doi/10.1103/PhysRevB.59.1758},
  year = {1999},
  type = {Journal Article}
}

@article{PerdewBurkeErnzerhof1996,
  author = {Perdew, John P. and Burke, Kieron and Ernzerhof, Matthias},
  title = {{G}eneralized {G}radient {A}pproximation {M}ade {S}imple},
  journal = {Phys. Rev. Lett.},
  volume = {77},
  pages = {3865--3868},
  number = {0},
  DOI = {10.1103/PhysRevLett.77.3865},
  url = {https://link.aps.org/doi/10.1103/PhysRevLett.77.3865},
  year = {1996},
  type = {Journal Article}
}

@article{Dudarev1998,
	title = {Electron-energy-loss spectra and the structural stability of nickel oxide:  An LSDA+U study},
	author = {Dudarev, S. L. and Botton, G. A. and Savrasov, S. Y. and Humphreys, C. J. and Sutton, A. P.},
	journal = {Phys. Rev. B},
	volume = {57},
	issue = {3},
	pages = {1505--1509},
	numpages = {0},
	year = {1998},
	month = {Jan},
	publisher = {American Physical Society},
	doi = {10.1103/PhysRevB.57.1505},
	url = {https://link.aps.org/doi/10.1103/PhysRevB.57.1505}
}

@article{LiangRuO2U2023,
	title = {First-principles study of the magnetic exchange forces between the RuO2(110) surface and Fe tip},
	journal = {ChemPhysChem},
	volume = {24},
	number = {e202200429},
	pages = {1439-4235},
	year = {2023},
	doi = {https://doi.org/10.1002/cphc.202200429},
	url = {https://doi.org/10.1002/cphc.202200429},
	author = {Liang, Qiuhua and Brocks, Geert and Bieberle-Hütter, Anja}
}

@article{CaiNiOU2008,
	title = {Study on the ground state of NiO: The LSDA (GGA)+U method},
	journal = {Physica B: Condensed Matter},
	volume = {404},
	number = {1},
	pages = {89-94},
	year = {2009},
	issn = {0921-4526},
	doi = {https://doi.org/10.1016/j.physb.2008.10.009},
	url = {https://www.sciencedirect.com/science/article/pii/S092145260800447X},
	author = {Tuo Cai and Huilei Han and You Yu and Tao Gao and Jiguang Du and Lianghuan Hao}
}

@article{MorrowCa2NiOsO6,
author = {Morrow, Ryan and Samanta, Kartik and Saha Dasgupta, Tanusri and Xiong, Jie and Freeland, John W. and Haskel, Daniel and Woodward, Patrick M.},
title = {Magnetism in Ca2CoOsO6 and Ca2NiOsO6: Unraveling the Mystery of Superexchange Interactions between 3d and 5d Ions},
journal = {Chemistry of Materials},
volume = {28},
number = {11},
pages = {3666-3675},
year = {2016},
doi = {10.1021/acs.chemmater.6b00254},

URL = { 
    
        https://doi.org/10.1021/acs.chemmater.6b00254
    
    

},
eprint = { 
    
        https://doi.org/10.1021/acs.chemmater.6b00254
    
    

}

}

@article{MarzariVanderbilt1997,
	title = {Maximally localized generalized Wannier functions for composite energy bands},
	author = {Marzari, Nicola and Vanderbilt, David},
	journal = {Phys. Rev. B},
	volume = {56},
	issue = {20},
	pages = {12847--12865},
	numpages = {0},
	year = {1997},
	month = {Nov},
	publisher = {American Physical Society},
	doi = {10.1103/PhysRevB.56.12847},
	url = {https://link.aps.org/doi/10.1103/PhysRevB.56.12847}
}

@article{Mostofi2014,
	title = {An updated version of wannier90: A tool for obtaining maximally-localised Wannier functions},
	journal = {Computer Physics Communications},
	volume = {185},
	number = {8},
	pages = {2309-2310},
	year = {2014},
	issn = {0010-4655},
	doi = {https://doi.org/10.1016/j.cpc.2014.05.003},
	url = {https://www.sciencedirect.com/science/article/pii/S001046551400157X},
	author = {Arash A. Mostofi and Jonathan R. Yates and Giovanni Pizzi and Young-Su Lee and Ivo Souza and David Vanderbilt and Nicola Marzari}
}

@article{MarzariReview2012,
	title = {Maximally localized {W}annier functions: {T}heory and applications},
	author = {Marzari, Nicola and Mostofi, Arash A. and Yates, Jonathan R. and Souza, Ivo and Vanderbilt, David},
	journal = {Rev. Mod. Phys.},
	volume = {84},
	issue = {4},
	pages = {1419--1475},
	numpages = {0},
	year = {2012},
	month = {Oct},
	publisher = {American Physical Society},
	doi = {10.1103/RevModPhys.84.1419},
	url = {https://link.aps.org/doi/10.1103/RevModPhys.84.1419}
}

@article{WuWannierTools2018,
	title = {Wannier{T}ools: An open-source software package for novel topological materials},
	journal = {Computer Physics Communications},
	volume = {224},
	pages = {405-416},
	year = {2018},
	issn = {0010-4655},
	doi = {https://doi.org/10.1016/j.cpc.2017.09.033},
	url = {https://www.sciencedirect.com/science/article/pii/S0010465517303442},
	author = {QuanSheng Wu and ShengNan Zhang and Hai-Feng Song and Matthias Troyer and Alexey A. Soluyanov}
}

@misc{YueWannhrSymmMag,
	author = {Yue, Changming},
	title  = {\texttt{wannhr\_symm\_Mag}: A tool for symmetrization of magnetic Wannier tight-binding Hamiltonians},
	note   = {\url{https://github.com/quanshengwu/wannier_tools/tree/master/utility/wannhr_symm_Mag}}
}

@article{Jiang2025DwaveKV2Se2O,
  author = {Jiang, Bei and Hu, Mingzhe and Bai, Jianli and Song, Ziyin and Mu, Chao and Qu, Gexing and Li, Wan and Zhu, Wenliang and Pi, Hanqi and Wei, Zhongxu and Sun, Yu-Jie and Huang, Yaobo and Zheng, Xiquan and Peng, Yingying and He, Lunhua and Li, Shiliang and Luo, Jianlin and Li, Zheng and Chen, Genfu and Li, Hang and Weng, Hongming and Qian, Tian},
  title = {{A metallic room-temperature {d}-wave altermagnet}},
  journal = {Nature Physics},
  volume = {21},
  number = {5},
  pages = {754--759},
  year = {2025},
  doi = {10.1038/s41567-025-02822-y},
}

@article{Nagaosa2010, author = {Nagaosa, Naoto and Sinova, Jairo and Onoda, Shigeki and MacDonald, A. H. and Ong, N. P.}, title = {Anomalous Hall effect}, journal= {Rev. Mod. Phys.}, volume = {82}, pages = {1539--1592}, year = {2010}, doi = {10.1103/RevModPhys.82.1539} }

@article{RevModPhys.95.011002,
  title = {Colloquium: Quantum anomalous Hall effect},
  author = {Chang, Cui-Zu and Liu, Chao-Xing and MacDonald, Allan H.},
  journal = {Rev. Mod. Phys.},
  volume = {95},
  issue = {1},
  pages = {011002},
  numpages = {33},
  year = {2023},
  month = {Jan},
  publisher = {American Physical Society},
  doi = {10.1103/RevModPhys.95.011002},
  url = {https://link.aps.org/doi/10.1103/RevModPhys.95.011002}
}

@article{Sinova2015, author = {Sinova, Jairo and Valenzuela, S. O. and Wunderlich, J. and Back, C. H. and Jungwirth, T.}, title = {Spin Hall effects}, journal= {Rev. Mod. Phys.}, volume = {87}, pages = {1213--1259}, year = {2015}, doi = {10.1103/RevModPhys.87.1213} }

@article{Liu2018, author = {Liu, Zhao and Zhao, Gan and Liu, Bing and Wang, Z. F. and Yang, Jinlong and Liu, Feng}, title = {Intrinsic quantum anomalous Hall effect with in-plane magnetization: Searching rule and material prediction}, journal= {Phys. Rev. Lett.}, volume = {121}, pages = {246401}, year = {2018}, doi = {10.1103/PhysRevLett.121.246401} }

@article{Sun2022, author = {Sun, Song and Weng, Hongming and Dai, Xi}, title = {Possible quantization and half-quantization in the anomalous Hall effect caused by in-plane magnetic field}, journal= {Phys. Rev. B}, volume = {106}, pages = {L241105}, year = {2022}, doi = {10.1103/PhysRevB.106.L241105} }

@article{Cao2023, author = {Cao, Jin and Jiang, Wei and Li, Xiao-Ping and Tu, Daifeng and Zhou, Jiadong and Zhou, Jianhui and Yao, Yugui}, title = {In-plane anomalous Hall effect in 
P
T
PT-symmetric antiferromagnetic materials}, journal= {Phys. Rev. Lett.}, volume = {130}, pages = {166702}, year = {2023}, doi = {10.1103/PhysRevLett.130.166702} }

@article{Battilomo2021, author = {Battilomo, Raffaele and Scopigno, Niccol`o and Ortix, Carmine}, title = {Anomalous planar Hall effect in two-dimensional trigonal crystals}, journal= {Phys. Rev. Research}, volume = {3}, pages = {L012006}, year = {2021}, doi = {10.1103/PhysRevResearch.3.L012006} }

@article{Lee2025EuZn2Sb2,
  title = {Difference in the in-plane anomalous Hall response in thin films of the Zintl compound $\mathrm{Eu}{A}_{2}{\mathrm{Sb}}_{2}$ ($A=\mathrm{Zn},\mathrm{Cd}$)},
  author = {Lee, Hsiang and Nishihaya, Shinichi and Kriener, Markus and Fujioka, Jun and Nakamura, Ayano and Watanabe, Yuto and Ishizuka, Hiroaki and Uchida, Masaki},
  journal = {Phys. Rev. B},
  volume = {111},
  issue = {24},
  pages = {L241106},
  numpages = {6},
  year = {2025},
  month = {Jun},
  publisher = {American Physical Society},
  doi = {10.1103/PhysRevB.111.L241106},
  url = {https://link.aps.org/doi/10.1103/PhysRevB.111.L241106}
}

@article{Wang2025Co3Sn2S2,
  title = {In-plane Hall effect in ${\mathrm{Co}}_{3}{\mathrm{Sn}}_{2}{\mathrm{S}}_{2}$},
  author = {Wang, Lujunyu and Tian, Guang and Chen, Haiyun and Zhao, Jiaji and Shi, Hengjie and Wang, Haoyu and Li, Zhilin and Wu, Xiaosong},
  journal = {Phys. Rev. B},
  volume = {111},
  issue = {5},
  pages = {054412},
  numpages = {7},
  year = {2025},
  month = {Feb},
  publisher = {American Physical Society},
  doi = {10.1103/PhysRevB.111.054412},
  url = {https://link.aps.org/doi/10.1103/PhysRevB.111.054412}
}

@article{Dai2026, author = {Dai, Kunjie and Wang, Zhen and Wu, Wenfeng and Jin, Feng and Hua, Enda and Liu, Nan and Lu, Jingdi and Zhang, Jinfeng and Zhao, Yuyue and Yang, Linda and Liu, Kai and Ye, Huan and Lv, Qiming and Liang, Zhengguo and Wang, Ao and Hou, Dazhi and Gao, Yang and Shen, Shengchun and Tao, Jing and Si, Liang and Wu, Wenbin and Wang, Lingfei}, title = {Symmetry-engineered and electrically tunable in-plane anomalous Hall effect in oxide heterostructures}, journal= {arXiv:2601.05462}, year = {2026},url={https://arxiv.org/abs/2601.05462} }

@article{PhysRevB.100.165117,
  title = {In-plane magnetic-field-induced quantum anomalous Hall plateau transition},
  author = {Zhang, Jinlong and Liu, Zhaochen and Wang, Jing},
  journal = {Phys. Rev. B},
  volume = {100},
  issue = {16},
  pages = {165117},
  numpages = {8},
  year = {2019},
  month = {Oct},
  publisher = {American Physical Society},
  doi = {10.1103/PhysRevB.100.165117},
  url = {https://link.aps.org/doi/10.1103/PhysRevB.100.165117}
}

@article{PhysRevB.72.104430,
  title = {Antisymmetric contribution to the planar Hall effect of ${\mathrm{Fe}}_{3}\mathrm{Si}$ films grown on $\mathrm{GaAs}(113)A$ substrates},
  author = {Muduli, P. K. and Friedland, K.-J. and Herfort, J. and Sch\"onherr, H.-P. and Ploog, K. H.},
  journal = {Phys. Rev. B},
  volume = {72},
  issue = {10},
  pages = {104430},
  numpages = {9},
  year = {2005},
  month = {Sep},
  publisher = {American Physical Society},
  doi = {10.1103/PhysRevB.72.104430},
  url = {https://link.aps.org/doi/10.1103/PhysRevB.72.104430}
}

@article{Chen2025, author = {Chen, Jiaxin and Zheng, Hongsheng and Chen, Hongliang and Shen, Qia and Pan, Chang and Zheng, Zhenyi and Yi, Hemian and Guan, Dandan and Liu, Xiaoxue and Li, Yaoyi and Wang, Shiyong and Zheng, Hao and Liu, Canhua and Jia, Jinfeng and Chen, Jingsheng and Zhong, Ruidan and Wang, Lei and Qiu, Xuepeng and Yang, Yumeng and Manchon, Aur'elien and Liu, Liang}, title = {Coexistence of unconventional spin-orbit torque and in-plane Hall effect in a single ferromagnetic layer}, journal= {arXiv:2511.17231}, year = {2025}, url={https://arxiv.org/abs/2511.17231} }

@article{rv1n-vr4p,
  title = {Third-Order Nonlinear Hall Effect in Altermagnet ${\mathrm{RuO}}_{2}$},
  author = {Chu, R. Y. and Han, L. and Gong, Z. H. and Fu, X. Z. and Bai, H. and Liang, S. X. and Chen, C. and Cheong, S-W. and Zhang, Y. Y. and Liu, J. W. and Wang, Y. Y. and Pan, F. and Lu, H. Z. and Song, C.},
  journal = {Phys. Rev. Lett.},
  volume = {135},
  issue = {21},
  pages = {216703},
  numpages = {8},
  year = {2025},
  month = {Nov},
  publisher = {American Physical Society},
  doi = {10.1103/rv1n-vr4p},
  url = {https://link.aps.org/doi/10.1103/rv1n-vr4p}
}

@article{Zhang2025SpinValleyRbV2Te2O,
  author = {Zhang, Fayuan and Cheng, Xingkai and Yin, Zhouyi and
            Liu, Changchao and Deng, Liwei and Qiao, Yuxi and
            Shi, Zheng and Zhang, Shuxuan and Lin, Junhao and
            Liu, Zhengtai and Ye, Mao and Huang, Yaobo and
            Meng, Xiangyu and Zhang, Cheng and Okuda, Taichi and
            Shimada, Kenya and Cui, Shengtao and Zhao, Yue and
            Cao, Guang-Han and Qiao, Shan and Liu, Junwei and
            Chen, Chaoyu},
  title = {Crystal-symmetry-paired spin--valley locking in a
           layered room-temperature metallic altermagnet candidate},
  journal = {Nature Physics},
  volume = {21},
  number = {5},
  pages = {760--767},
  year = {2025},
  doi = {10.1038/s41567-025-02864-2},
  url = {https://doi.org/10.1038/s41567-025-02864-2}
}

@article{Hu2026SpinValleyNodalLines,
  author = {Hu, Quanxin and Cheng, Xingkai and Duan, Qingchen and
            Hu, Yudong and Jiang, Bei and Xiao, Yusen and
            Li, Yaqi and Pan, Mojun and Deng, Liwei and
            Liu, Changchao and Cao, Guanghan and Liu, Zhengtai and
            Ye, Mao and Qiao, Shan and Liu, Zhanfeng and
            Sun, Zhe and Gao, Anyuan and Huang, Yaobo and
            Zhong, Ruidan and Liu, Junwei and Lv, Baiqing and
            Ding, Hong},
  title = {Observation of spin-valley locked nodal lines in a
           quasi-{2D} altermagnet},
  year = {2026},
  journal = {arXiv:2601.02883},
  url = {https://arxiv.org/abs/2601.02883}
}

@article{Fedchenko2024RuO2,
  author  = {Fedchenko, Olena and Min{\'a}r, Jan and Akashdeep, Akashdeep and D'Souza, Sunil Wilfred and Vasilyev, Dmitry and Tkach, Olena and Odenbreit, Lukas and Nguyen, Quynh and Kutnyakhov, Dmytro and Wind, Nils and Wenthaus, Lukas and Scholz, Markus and Rossnagel, Kai and Hoesch, Moritz and Aeschlimann, Martin and Stadtm{\"u}ller, Benjamin and Kl{\"a}ui, Mathias and Sch{\"o}nhense, Gerd and Jungwirth, Tomas and Hellenes, Anna Birk and Jakob, Gerhard and {\v{S}}mejkal, Libor and Sinova, Jairo and Elmers, Hans-Joachim},
  title   = {Observation of time-reversal symmetry breaking in the band structure of altermagnetic {RuO$_2$}},
  journal = {Science Advances},
  year    = {2024},
  volume  = {10},
  number  = {5},
  pages   = {eadj4883},
  doi     = {10.1126/sciadv.adj4883}
}

@article{PhysRevB.107.115142,
  title = {Higher-order nonlinear anomalous Hall effects induced by Berry curvature multipoles},
  author = {Zhang, Cheng-Ping and Gao, Xue-Jian and Xie, Ying-Ming and Po, Hoi Chun and Law, K. T.},
  journal = {Phys. Rev. B},
  volume = {107},
  issue = {11},
  pages = {115142},
  numpages = {16},
  year = {2023},
  month = {Mar},
  publisher = {American Physical Society},
  doi = {10.1103/PhysRevB.107.115142},
  url = {https://link.aps.org/doi/10.1103/PhysRevB.107.115142}
}

@article{ifmmode2022,
  title = {Emerging Research Landscape of Altermagnetism},
  author = {\ifmmode \check{S}\else \v{S}\fi{}mejkal, Libor and Sinova, Jairo and Jungwirth, Tomas},
  journal = {Phys. Rev. X},
  volume = {12},
  issue = {4},
  pages = {040501},
  numpages = {27},
  year = {2022},
  month = {Dec},
  publisher = {American Physical Society},
  doi = {10.1103/PhysRevX.12.040501},
  url = {https://link.aps.org/doi/10.1103/PhysRevX.12.040501}
}

@article{yrs7-m6zy,
  title = {Probing $\boldsymbol{k}$-Space Alternating Spin Polarization via the Anomalous Hall Effect},
  author = {Chen, Rui and Wang, Zi-Ming and Wu, Ke and Sun, Hai-Peng and Zhou, Bin and Wang, Rui and Xu, Dong-Hui},
  journal = {Phys. Rev. Lett.},
  volume = {135},
  issue = {9},
  pages = {096602},
  numpages = {8},
  year = {2025},
  month = {Aug},
  publisher = {American Physical Society},
  doi = {10.1103/yrs7-m6zy},
  url = {https://link.aps.org/doi/10.1103/yrs7-m6zy}
}

@ARTICLE{2025arXiv251005512L,
       author = {{Luo}, Xun-Jiang and {Hu}, Jin-Xin and {Hu}, Meng-Li and {Law}, K.~T.},
        title = "{Spin Group Symmetry Criteria for Odd-parity Magnets}",
      journal = {arXiv:2510.05512},
         year = 2025,
        month = oct,
       url={https://arxiv.org/abs/2510.05512}
}

@article{CheXiaobing2024,
  title = {Enumeration and Representation Theory of Spin Space Groups},
  author = {Chen, Xiaobing and Ren, Jun and Zhu, Yanzhou and Yu, Yutong and Zhang, Ao and Liu, Pengfei and Li, Jiayu and Liu, Yuntian and Li, Caiheng and Liu, Qihang},
  journal = {Phys. Rev. X},
  volume = {14},
  issue = {3},
  pages = {031038},
  numpages = {33},
  year = {2024},
  month = {Aug},
  publisher = {American Physical Society},
  doi = {10.1103/PhysRevX.14.031038},
  url = {https://link.aps.org/doi/10.1103/PhysRevX.14.031038}
}

@article{LiuPengfei2022,
  title = {Spin-Group Symmetry in Magnetic Materials with Negligible Spin-Orbit Coupling},
  author = {Liu, Pengfei and Li, Jiayu and Han, Jingzhi and Wan, Xiangang and Liu, Qihang},
  journal = {Phys. Rev. X},
  volume = {12},
  issue = {2},
  pages = {021016},
  numpages = {19},
  year = {2022},
  month = {Apr},
  publisher = {American Physical Society},
  doi = {10.1103/PhysRevX.12.021016},
  url = {https://link.aps.org/doi/10.1103/PhysRevX.12.021016}
}

@article{JiangYi2024,
  title = {Enumeration of Spin-Space Groups: Toward a Complete Description of Symmetries of Magnetic Orders},
  author = {Jiang, Yi and Song, Ziyin and Zhu, Tiannian and Fang, Zhong and Weng, Hongming and Liu, Zheng-Xin and Yang, Jian and Fang, Chen},
  journal = {Phys. Rev. X},
  volume = {14},
  issue = {3},
  pages = {031039},
  numpages = {25},
  year = {2024},
  month = {Aug},
  publisher = {American Physical Society},
  doi = {10.1103/PhysRevX.14.031039},
  url = {https://link.aps.org/doi/10.1103/PhysRevX.14.031039}
}

@article{XiaoZhenyu2024,
  title = {Spin Space Groups: Full Classification and Applications},
  author = {Xiao, Zhenyu and Zhao, Jianzhou and Li, Yanqi and Shindou, Ryuichi and Song, Zhi-Da},
  journal = {Phys. Rev. X},
  volume = {14},
  issue = {3},
  pages = {031037},
  numpages = {33},
  year = {2024},
  month = {Aug},
  publisher = {American Physical Society},
  doi = {10.1103/PhysRevX.14.031037},
  url = {https://link.aps.org/doi/10.1103/PhysRevX.14.031037}
}

@Article{Liu2026a,
author={Liu, Yuntian
and Chen, Xiaobing
and Yu, Yutong
and Etxebarria, Jes{\'u}s
and Perez-Mato, J. Manuel
and Liu, Qihang},
title={Symmetry classification of magnetic orders using oriented spin space groups},
journal={Nature},
year={2026},
month={Apr},
day={01},
volume={652},
number={8111},
pages={869-873},
issn={1476-4687},
doi={10.1038/s41586-026-10401-1},
url={https://doi.org/10.1038/s41586-026-10401-1}
}

@article{supp,
author = {},
journal = {See supplemental material for magnetic geometry described by  (SSG);
 magnetic geometry described by magnetic space group and symmetry condition for spontaneous IPAHE; field-induced IPAH; field direction dependence of Hall conductance; candidate materials for IPAH; details of first-principle calculations; effective two-band model for field-induced IPAHE; arbitrary-direction magnetic fields. },
title = {},
}

@Article{Liu2025a,
author={Liu, Qihang
and Dai, Xi
and Bl{\"u}gel, Stefan},
title={Different facets of unconventional magnetism},
journal={Nature Physics},
year={2025},
month={Mar},
day={01},
volume={21},
number={3},
pages={329-331},
issn={1745-2481},
doi={10.1038/s41567-024-02750-3},
url={https://doi.org/10.1038/s41567-024-02750-3}
}

@article{BaiLing2024,
author = {Bai, Ling and Feng, Wanxiang and Liu, Siyuan and Šmejkal, Libor and Mokrousov, Yuriy and Yao, Yugui},
title = {Altermagnetism: Exploring New Frontiers in Magnetism and Spintronics},
journal = {Advanced Functional Materials},
volume = {34},
number = {49},
pages = {2409327},
doi = {https://doi.org/10.1002/adfm.202409327},
url = {https://advanced.onlinelibrary.wiley.com/doi/abs/10.1002/adfm.202409327},
year = {2024}
}

@ARTICLE{BirkHellenes2023,
       author = {{Birk Hellenes}, Anna and {Jungwirth}, Tom{\'a}{\v{s}} and {Jaeschke-Ubiergo}, Rodrigo and {Chakraborty}, Atasi and {Sinova}, Jairo and {{\v{S}}mejkal}, Libor},
        title = "{P-wave magnets}",
      journal = {arXiv:2309.01607},
     url={https://arxiv.org/abs/2309.01607}
}

@ARTICLE{2026arXiv260307643L,
	       author = {{Luo}, Xun-Jiang and {Hu}, Jin-Xin and {Hu}, Mengli and {Law}, K.~T.},
	        title = "{Spin Group Symmetry Criteria For Unconventional Magnetism}",
	      journal = {arXiv:2603.07643},
	         year = 2026,
	     url={https://arxiv.org/abs/2603.07643}
	}

@ARTICLE{2026arXiv260521336L,
       author = {{Luo}, Xun-Jiang and {Li}, Dan and {Xiao}, Rui-Chun and {Shao}, Ding-Fu and {Li}, Lei and {Tian}, Mingliang and {Yao}, Yugui},
        title = "{Unconventional Magnetism: Symmetry Classification, Hybrid-parity and Unconstrained-parity Classes}",
      journal = {arXiv:2605.21336},
   url={https://arxiv.org/abs/2605.21336}
}

@article{PhysRevB.84.085123,
  title = {Quantized anomalous Hall insulator in a nanopatterned two-dimensional electron gas},
  author = {Zhang, Yongping and Zhang, Chuanwei},
  journal = {Phys. Rev. B},
  volume = {84},
  issue = {8},
  pages = {085123},
  numpages = {5},
  year = {2011},
  month = {Aug},
  publisher = {American Physical Society},
  doi = {10.1103/PhysRevB.84.085123},
  url = {https://link.aps.org/doi/10.1103/PhysRevB.84.085123}
}

@article{qiaozhenghua2016,
  title = {Quantum anomalous Hall effect in atomic crystal layers from in-plane magnetization},
  author = {Ren, Yafei and Zeng, Junjie and Deng, Xinzhou and Yang, Fei and Pan, Hui and Qiao, Zhenhua},
  journal = {Phys. Rev. B},
  volume = {94},
  issue = {8},
  pages = {085411},
  numpages = {5},
  year = {2016},
  month = {Aug},
  publisher = {American Physical Society},
  doi = {10.1103/PhysRevB.94.085411},
  url = {https://link.aps.org/doi/10.1103/PhysRevB.94.085411}
}

@Article{Zhou2022,
author={Zhou, Jiadong
and Zhang, Wenjie
and Lin, Yung-Chang
and Cao, Jin
and Zhou, Yao
and Jiang, Wei
and Du, Huifang
and Tang, Bijun
and Shi, Jia
and Jiang, Bingyan
and Cao, Xun
and Lin, Bo
and Fu, Qundong
and Zhu, Chao
and Guo, Wei
and Huang, Yizhong
and Yao, Yuan
and Parkin, Stuart S. P.
and Zhou, Jianhui
and Gao, Yanfeng
and Wang, Yeliang
and Hou, Yanglong
and Yao, Yugui
and Suenaga, Kazu
and Wu, Xiaosong
and Liu, Zheng},
title={Heterodimensional superlattice with in-plane anomalous Hall effect},
journal={Nature},
year={2022},
month={Sep},
day={01},
volume={609},
number={7925},
pages={46-51},
issn={1476-4687},
doi={10.1038/s41586-022-05031-2},
url={https://doi.org/10.1038/s41586-022-05031-2}
}

@article{PhysRevB.109.155153,
  title = {Switchable in-plane anomalous Hall effect by magnetization orientation in monolayer ${\mathrm{Mn}}_{3}{\mathrm{Si}}_{2}{\mathrm{Te}}_{6}$},
  author = {Li, Ding and Wang, Maoyuan and Li, Dengfeng and Zhou, Jianhui},
  journal = {Phys. Rev. B},
  volume = {109},
  issue = {15},
  pages = {155153},
  numpages = {8},
  year = {2024},
  month = {Apr},
  publisher = {American Physical Society},
  doi = {10.1103/PhysRevB.109.155153},
  url = {https://link.aps.org/doi/10.1103/PhysRevB.109.155153}
}

@article{PhysRevLett.111.086802,
  title = {In-Plane Magnetization-Induced Quantum Anomalous Hall Effect},
  author = {Liu, Xin and Hsu, Hsiu-Chuan and Liu, Chao-Xing},
  journal = {Phys. Rev. Lett.},
  volume = {111},
  issue = {8},
  pages = {086802},
  numpages = {5},
  year = {2013},
  month = {Aug},
  publisher = {American Physical Society},
  doi = {10.1103/PhysRevLett.111.086802},
  url = {https://link.aps.org/doi/10.1103/PhysRevLett.111.086802}
}

@article{PhysRevB.103.214438,
  title = {Unconventional anomalous Hall effect from magnetization parallel to the electric field},
  author = {Tan, Hengxin and Liu, Yizhou and Yan, Binghai},
  journal = {Phys. Rev. B},
  volume = {103},
  issue = {21},
  pages = {214438},
  numpages = {5},
  year = {2021},
  month = {Jun},
  publisher = {American Physical Society},
  doi = {10.1103/PhysRevB.103.214438},
  url = {https://link.aps.org/doi/10.1103/PhysRevB.103.214438}
}

@article{liu2025multipolar,
  author  = {Liu, Z. and Wei, M. and Peng, W. and Hou, D. and Gao, Y. and Niu, Q.},
  title   = {Multipolar anisotropy in anomalous {Hall} effect from spin-group symmetry breaking},
  journal = {Phys. Rev. X},
  volume  = {15},
  pages   = {031006},
  year    = {2025},
  doi     = {10.1103/PhysRevX.15.031006},
  note    = {arXiv:2408.08810}
}

@article{peng2026octupole,
  author  = {Peng, W. and Liu, Z. and Pan, H. and Wang, P. and Chen, Y. and Zhang, J. and Yu, X. and Shen, J. and Yang, M. and Niu, Q. and Gao, Y. and Hou, D.},
  title   = {Uncovering the magnetization multipolar anisotropy of the anomalous {Hall} effect in conventional ferromagnets},
  journal = {Chin. Phys. Lett.},
  volume  = {43},
  pages   = {060704},
  year    = {2026},
  doi     = {10.1088/0256-307X/43/6/060704}
}

@article{chen2025strain,
  author  = {Chen, H. and Cui, Y. and Wu, T. and others and Xiao, C. and Wu, Y.},
  title   = {Strain-induced in-plane anomalous {Hall} effect in single-crystalline {Fe}(211) films},
  journal = {Phys. Rev. B},
  volume  = {111},
  pages   = {174423},
  year    = {2025},
  doi     = {10.1103/PhysRevB.111.174423}
}

@article{nishihaya2025sro,
  author  = {Nishihaya, S. and Matsuki, Y. and Kaminakamura, H. and Murakami, Y. and Arita, R. and Ishizuka, H. and Uchida, M.},
  title   = {Spontaneous in-plane anomalous {Hall} response observed in a ferromagnetic oxide},
  journal = {Adv. Mater.},
  volume  = {37},
  pages   = {2502624},
  year    = {2025},
  doi     = {10.1002/adma.202502624}
}

@article{sankar2025fe3sn,
  author  = {Sankar, Soumya and Cheng, Xingkai and Murtaza, Tahir and Chen, Caiyun and Qin, Yuqi and Wu, Xuezhao and Shao, Qiming and Lortz, Rolf and Liu, Junwei and J{\"a}ck, Berthold},
  title   = {Room temperature observation of the anomalous in-plane {Hall} effect in a {Weyl} ferromagnet},
  journal = {Nat. Commun.},
  volume  = {17},
  pages   = {423},
  year    = {2026},
  doi     = {10.1038/s41467-025-67111-x}
}

@Article{kao2026lowdim,
author={Kao, I-Hsuan
and Bandapelli, Ravi Kumar
and Cui, Zhenhong
and Zhang, Shuchen
and Tang, Jian
and Qian, Tiema
and Sasmal, Souvik
and Tiwari, Aalok
and Chen, Mei-Tung
and Posti, Raghvendra
and Rao, Rahul
and Li, Jiahan
and Edgar, James H.
and Watanabe, Kenji
and Taniguchi, Takashi
and Ni, Ni
and Xu, Su-Yang
and Ma, Qiong
and Chatterjee, Shubhayu
and Katoch, Jyoti
and Singh, Simranjeet},
title={In-plane anomalous Hall effect in a low-dimensional system},
journal={Nature Materials},
year={2026},
url={https://doi.org/10.1038/s41563-026-02611-9}
}

@article{zheng2026cr3te4,
  author        = {Zheng, G. and Bordoloi, A. and Fan, M. and Kitou, S. and Saito, H. and Nakajima, T. and Singh, S. and Kurumaji, T. and Ye, L.},
  title         = {Dominant in-plane anomalous {Hall} effect in a monoclinic room-temperature ferromagnet},
  journal={arXiv:2606.10063},
  year          = {2026},
  url={https://arxiv.org/abs/2606.10063}
}

@misc{peng2026cr12te2,
  author        = {Peng, W. and Liu, Z. and Wang, S. and Pan, H. and others and Xiang, B. and Hou, D.},
  title         = {Pronounced in-plane anomalous {Hall} effect with vanishing out-of-plane response in {Cr}$_{1.2}${Te}$_2$},
  year          = {2026},
  eprint        = {2606.14272},
  archivePrefix = {arXiv}
}

@article{Miao2024Cd3As2,
  author  = {Miao, W. and Guo, B. and Stemmer, S. and Dai, X.},
  title   = {Engineering the in-plane anomalous {Hall} effect in {Cd}$_3${As}$_2$ thin films},
  journal = {Phys. Rev. B},
  volume  = {109},
  pages   = {155408},
  year    = {2024},
  doi     = {10.1103/PhysRevB.109.155408}
}

@article{Nakamura2024,
  author  = {Nakamura, A. and Nishihaya, S. and Ishizuka, H. and Kriener, M. and Watanabe, Y. and Uchida, M.},
  title   = {In-plane anomalous {Hall} effect associated with orbital magnetization: Measurements of low-carrier-density films of a magnetic {Weyl} semimetal},
  journal = {Phys. Rev. Lett.},
  volume  = {133},
  pages   = {236602},
  year    = {2024},
  doi     = {10.1103/PhysRevLett.133.236602}
}

\clearpage

\begin{widetext}
\begin{center}
\begin{large}
\textbf{Supplemental Material for ‘‘A Unified Symmetry Framework For In-plane Anomalous Hall effect"}
\end{large}
\end{center}

\setcounter{figure}{0}
\setcounter{equation}{0}
\setcounter{table}{0}
\renewcommand\thefigure{S\arabic{figure}}
\renewcommand\thetable{S\arabic{table}}
\renewcommand\theequation{S\arabic{equation}}

This Supplemental Material includes the following eight sections:
(1) magnetic geometry described by the spin space group (SSG);
(2) magnetic geometry described by the magnetic space group and the symmetry condition for the spontaneous in-plane anomalous Hall effect (IPAHE);
(3) field-induced IPAHE;
(4) field direction dependence of Hall conductance;
(5) candidate materials for IPAHE;
(6) details of first-principles calculations;
(7) effective two-band model for field-induced IPAHE;
(8) IPAHE under arbitrary-direction magnetic fields.

\section{A. Magnetic geometry described by SSG}
\label{app:spontaneous_IPAHE}

The SSG fully determines the magnetic geometry of a material. In Ref.~\cite{Liu2026a}, the authors established the symmetry criteria for ferromagnets and antiferromagnets by extracting the spin operations from the SSG, which form a point group $\mathcal{P}_s$. Here, we demonstrate that this point group also encodes the symmetry-allowed directions of the net magnetization for ferromagnetic states.

A general SSG operation is denoted by $g_s = \{U||R\}$, where $R \in O(3)$ acts on the spatial coordinates and $U \in O(3)$ acts independently on the spin degrees of freedom. The set of all spin operations $\{U\}$ obtained by projecting the SSG onto the spin sector forms a finite subgroup $\mathcal{P}_s \subseteq O(3)$. That $\mathcal{P}_s$ is indeed a group follows from the group structure of the SSG itself: it contains the identity $\{\mathbb{I}||\mathbb{I}\}$; the product of any two SSG operations, $\{U_1||R_1\}\{U_2||R_2\} = \{U_1U_2||R_1R_2\}$, remains an SSG operation; and every element has an inverse. Consequently, the spin-sector projection inherits associativity, closure under multiplication, and the existence of inverses, making $\mathcal{P}_s$ a well-defined group.

A net magnetization $\boldsymbol{M}$ can exist in a magnetic structure if and only if $\mathcal{P}_s$ is polar, meaning that there exists a common invariant direction for all $U \in \mathcal{P}_s$. The space of all such allowed magnetization directions is the {spin polarization space}
\begin{equation}
\mathcal{V}_s = \bigcap_{U \in \mathcal{P}_s} \ker(U - \mathbb{I}),
\label{eq:Vs}
\end{equation}
where $\mathbb{I}$ is the $3 \times 3$ identity matrix and $\ker(U - \mathbb{I})$ denotes the eigenspace of $U$ with eigenvalue $+1$, i.e., the spin directions left invariant by $U$. Physically, any net magnetization $\boldsymbol{M}$ must lie strictly within $\mathcal{V}_s$. For example, consider a collinear ferromagnet with the net magnetization along the $z$ direction. The spin polarization space $\mathcal{V}_s$ is one-dimensional and is spanned by the vector $(0,0,1)$, an eigenvector of the preserved symmetry $U_z(\theta)$, i.e., an arbitrary rotation about the $z$ axis. In this case, $\dim(\mathcal{V}_s)=1$. 

More generally, the dimensionality of $\mathcal{V}_s$ reflects the magnetic order and the underlying spin symmetry. For a collinear ferromagnet, where all spins are aligned along a single axis, $\dim(\mathcal{V}_s)=1$. For a coplanar magnetic structure, where all spins lie within a common plane, the dimensionality can be either $1$ or $2$. The case $\dim(\mathcal{V}_s)=1$ corresponds to a net magnetization along a single direction within the plane, with any perpendicular in-plane component canceled out, as realized in certain canted antiferromagnets. The case $\dim(\mathcal{V}_s)=2$ requires that the net magnetization be allowed along any direction within the common spin plane. Similarly, for a noncoplanar magnetic structure, $\dim(\mathcal{V}_s)$ can be $1$, $2$, or $3$, depending on whether the point group $\mathcal{P}_s$ fixes an axis, a plane, or no direction at all.

Notably, even for a fully compensated magnetic structure with zero net magnetization, the spin polarization space $\mathcal{V}_s$ may remain nonempty if $\mathcal{P}_s$ is polar. In such cases, the symmetry permits a net magnetization along the invariant direction, but the specific magnetic configuration (e.g., antiparallel sublattices of equal magnitude) results in complete cancellation, yielding $\boldsymbol{M}=0$ while $\mathcal{V}_s \neq \{\boldsymbol{0}\}$.

\section{B. Magnetic geometry described by MSG and symmetry condition for spontaneous IPAHE}
\label{app:msg_IPAHE}

In this section, we present the magnetic geometry description under the MSG framework and detail the symmetry requirements for spontaneous IPAHE.

\subsection{B1. Magnetic geometry described by magnetic space group}

An antiferromagnet is defined as a magnetic structure with symmetry-enforced zero net magnetization. This enforcement can operate at the level of either the SSG or the MSG. Under the MSG framework, spin-orbit coupling (SOC) is activated, and the spin and spatial degrees of freedom are fully coupled. One can extract the effective spin-sector operations from the MSG (or, equivalently, the magnetic point group, MPG); these form a point group $\mathcal{P}_m \subseteq O(3)$. The net magnetization direction space is then defined by the fixed subspace of $\mathcal{P}_m$,
\begin{equation}
\mathcal{V}_m = \bigcap_{A \in \mathcal{P}_m} \ker(A - \mathbb{I}),
\label{eq:Vm}
\end{equation}
where $\mathbb{I}$ is the $3\times 3$ identity matrix. If $\mathcal{P}_m$ is nonpolar, so that $\dim(\mathcal{V}_m)=0$, the net magnetization is symmetry-forbidden, and the system is a pure antiferromagnet. If $\mathcal{P}_m$ is polar with $\dim(\mathcal{V}_m)\geq 1$, the net magnetization is symmetry-allowed, and the system is classified as a ferromagnet or ferrimagnet under the MSG framework. Among the 122 MPGs, 31 are ferromagnetic [$\dim(\mathcal{V}_m)\geq 1$], and the remaining 91 are non-ferromagnetic [$\dim(\mathcal{V}_m)=0$].

The SSG and MSG can be unified within the recently developed oriented SSG framework \cite{Liu2026a}, which is an SSG with a fixed magnetic orientation. In this unified picture, the relation between the MSG and the SSG is clear: the MSG is a subgroup of the SSG. Consequently, $\mathcal{P}_m \subseteq \mathcal{P}_s$, which leads to the fundamental subspace relation
\begin{equation}
\mathcal{V}_s \subseteq \mathcal{V}_m.
\end{equation}
This means that any magnetization allowed by the SSG symmetry is also allowed by the MSG symmetry. Comparing the magnetic geometries under the SSG and MSG frameworks reveals three distinct scenarios:
\begin{itemize}
\item Both the SSG and the MSG define an antiferromagnet; that is, both symmetries forbid a net magnetization ($\mathcal{P}_s$ and $\mathcal{P}_m$ are both nonpolar). This includes the well-known $P\mathcal{T}$-symmetric antiferromagnets.
\item Both the SSG and the MSG define a ferromagnet; that is, both symmetries allow a net magnetization ($\mathcal{P}_s$ and $\mathcal{P}_m$ are both polar). This describes conventional collinear and noncollinear ferromagnets.
\item The SSG defines an antiferromagnet but the MSG defines a ferromagnet. Here, $\mathcal{P}_s$ is nonpolar while $\mathcal{P}_m$ is polar, corresponding to $\dim(\mathcal{V}_s)=0$ and $\dim(\mathcal{V}_m)\geq 1$. This scenario has recently been identified as spin-orbital magnetism and systematically screened from the MAGNDATA database \cite{Liu2026a}.
\end{itemize}

\subsection{B2. Symmetry condition for spontaneous IPAHE}

Under the MSG framework with SOC activated, real-space axial vectors satisfy the same symmetry constraints as the spin magnetization. Consequently, the anomalous Hall conductivity vector $\boldsymbol{\sigma}^{\mathrm{H}}$, defined by $(\boldsymbol{\sigma}^{\mathrm{H}})_i = \epsilon_{ijk}\sigma_{jk}/2$ with $\epsilon_{ijk}$ the Levi-Civita symbol and $\sigma_{jk}$ the Hall conductivity tensor, obeys the same symmetry properties as the net magnetization $\boldsymbol{M}$. Both $\boldsymbol{M}$ and $\boldsymbol{\sigma}^{\mathrm{H}}$ are axial vectors, and their symmetry-allowed directions are governed by the point group $\mathcal{P}_m$.

Conventionally, $\boldsymbol{\sigma}^{\mathrm{H}}$ is collinear with $\boldsymbol{M}$, corresponding to the conventional anomalous Hall effect, in which the applied electric field, the Hall current, and the magnetization are mutually orthogonal. In this context, the net magnetization $\boldsymbol{M}$ is assumed to be set by the SSG symmetry (constrained to $\mathcal{V}_s$), while $\boldsymbol{\sigma}^{\mathrm{H}}$ is set by the MSG symmetry (constrained to $\mathcal{V}_m$). The difference between the two frameworks naturally leads to the scenario in which SOC enlarges the polarization space by breaking the spin-group symmetry. When $\mathcal{V}_s$ is a proper subspace of $\mathcal{V}_m$,
\begin{equation}
\mathcal{V}_s \subsetneq \mathcal{V}_m, \quad \dim(\mathcal{V}_m) > \dim(\mathcal{V}_s) > 0,
\end{equation}
the Hall vector $\boldsymbol{\sigma}^{\mathrm{H}}$ can acquire a component orthogonal to the net magnetization $\boldsymbol{M}$ fixed by the SSG. The resulting Hall current then flows within the plane containing $\boldsymbol{M}$, giving rise to the IPAHE. Therefore, the symmetry condition for spontaneous IPAHE in a ferromagnet is $\dim(\mathcal{V}_m) > \dim(\mathcal{V}_s) > 0$. Spin-orbital magnetism, in contrast, corresponds to $\dim(\mathcal{V}_m) \geq 1$ and $\dim(\mathcal{V}_s)=0$. Both phenomena represent spin-group symmetry-breaking effects induced by SOC.

We emphasize that this symmetry condition for IPAHE is necessary but not sufficient. Microscopically, the time-reversal-odd anomalous Hall effect is given by the integral of the Berry curvature over the Brillouin zone, which must be evaluated on a case-by-case basis.

\section{C. Field-induced IPAHE}
\label{app:field_induced_ipah}

In this section, we detail the application of our symmetry framework for field-induced IPAHE to all 122 MPGs. We summarize the surviving symmetry operations and the reduced MPGs, $G_m'$, for each initial MPG $G_m$ under a magnetic field applied along the three orthogonal axes of the conventional Cartesian coordinate system. Based on $G_m$ and $G_m'$, we identify the MPGs that can support field-induced IPAHE.

\subsection{C1. Classification of MPGs and crystallographic axis conventions}

The 122 MPGs are classified into three types according to whether and how time-reversal symmetry $\mathcal{T}$ is incorporated. The 32 Type-I (colorless) groups correspond to the 32 conventional crystallographic point groups $G$, containing only unitary spatial operations without $\mathcal{T}$. The 32 Type-II (gray) groups are the direct products $G\times\{E,\mathcal{T}\}$, with $E$ being the identity operation; these describe nonmagnetic states and magnetic materials with time-reversal translation ($\mathcal{T}\tau$) symmetry. The 58 Type-III (black-and-white) groups are constructed from the index-2 normal subgroups $H$ of crystallographic point groups $G$ via $M=H+(G-H)\mathcal{T}$. Here, $H$ contains the unitary operations, while $(G-H)\mathcal{T}$ contains the antiunitary ones. Different choices of $H$ for the same $G$ yield distinct Type-III groups. These describe either ferromagnets or antiferromagnets depending on the choice of $H$: if there exists a common polarization direction preserved by $H$ but reversed by every operation in $(G-H)$, the resulting MPG is ferromagnetic; otherwise, it is antiferromagnetic.

The 122 MPGs are distributed across the seven crystal systems: 5 in triclinic, 11 in monoclinic, 12 in orthorhombic, 31 in tetragonal, 16 in trigonal, 31 in hexagonal, and 16 in cubic. In the symmetry analysis below, we adopt standard crystallographic axis conventions: the unique twofold axis in the monoclinic system is set along $y$ ($C_{2y}$); the three mutually perpendicular twofold axes (or mirror planes) in the orthorhombic system are aligned along $x$, $y$, and $z$; and the highest-order rotational axes in the tetragonal ($C_{4z}$), trigonal ($C_{3z}$), and hexagonal ($C_{6z}$) systems are all aligned along $z$. The triclinic system possesses no rotational axes, while the cubic system contains four threefold axes along the body diagonals, in addition to twofold or fourfold axes along the Cartesian directions.

Among the 122 MPGs, 31 are ferromagnetic, admitting a spontaneous magnetization along at least one symmetry-allowed direction. These comprise 13 Type-I groups---$1$, $\bar{1}$, $2$, $m$, $2/m$, $4$, $\bar{4}$, $4/m$, $3$, $\bar{3}$, $6$, $\bar{6}$, $6/m$---and 18 Type-III groups---$2'$, $m'$, $2'/m'$, $2'2'2$, $m'm2'$, $m'm'2$, $m'm'm$, $42'2'$, $4m'm'$, $\bar{4}2'm'$, $4/mm'm'$, $32'$, $3m'$, $\bar{3}m'$, $62'2'$, $6m'm'$, $\bar{6}m'2'$, and $6/mm'm'$. In these 31 ferromagnetic point groups, MPGs $1$ and $\bar{1}$ in the triclinic crystal system allow a net magnetization $\boldsymbol{M}$ along arbitrary directions. MPGs $2'$, $m'$, $2'/m'$ in the monoclinic crystal system allow $\boldsymbol{M}$ along the $x$ and $y$ directions. The remaining 26 ferromagnetic point groups allow a net magnetization only along the $z$ direction.

\subsection{C2. Surviving symmetries and reduced MPGs}

For an initial MPG $G_m$, we consider applying a magnetic field $\boldsymbol{B}$ along the direction $\boldsymbol{d}$. As an axial vector, $\boldsymbol{B}$ breaks time-reversal symmetry and lowers spatial symmetry, preserving only operations that leave $\boldsymbol{B}$ invariant. The surviving operations form the group
\begin{equation}
\mathcal{S}_{\boldsymbol{d}} = \{I, C_{n\boldsymbol{d}}, m_{\boldsymbol{d}}, m'_{\boldsymbol{p}}, C'_{2\boldsymbol{p}}\},
\end{equation}
where $I$ is inversion, $C_{n\boldsymbol{d}}$ is the $n$-fold rotation about $\boldsymbol{d}$, and $m_{\boldsymbol{d}}$ is the mirror plane perpendicular to $\boldsymbol{d}$. The antiunitary operations $m'_{\boldsymbol{p}}$ and $C'_{2\boldsymbol{p}}$ combine time reversal with the mirror plane perpendicular to $\boldsymbol{p}$ and the twofold rotation about $\boldsymbol{p}$ ($\boldsymbol{p} \perp \boldsymbol{d}$), respectively. The field-lowered MPG is the intersection of $G_m$ and $\mathcal{S}_{\boldsymbol{d}}$,
\begin{equation}
G_m' = G_m \cap \mathcal{S}_{\boldsymbol{d}},
\end{equation}
which constitutes the maximal common subgroup of $G_m$ and $\mathcal{S}_{\boldsymbol{d}}$.

Applying this framework to all 122 MPGs, we align the magnetic field along the three orthogonal Cartesian axes $\boldsymbol{d} = x, y, z$ and identify the surviving symmetry operations and reduced MPGs $G_m'$ for each initial MPG. The results are enumerated in Tables~\ref{tab:s1}, \ref{tab:s2}, and \ref{tab:s3}, respectively.

\subsection{C3. Symmetry condition for field-induced IPAHE}
 Under an applied field $\boldsymbol{B} \parallel \boldsymbol{d}$, the MPG is lowered from $G_m$ to $G_m'$. The field-lowered Hall response space $\mathcal{V}_m'$ is the invariant subspace of the point group $\mathcal{P}_m'$, the set of spin operations of $G_m'$:
\begin{equation}
\mathcal{V}_m' = \bigcap_{A \in \mathcal{P}_m'} \ker(A - \mathbb{I}) 
\end{equation}
The constrained polarization space $\mathcal{V}_s'$ is defined by combining the initially allowed anomalous Hall response space $\mathcal{V}_m$ (determined by $G_m$) with the field direction $\boldsymbol{d}$:
\begin{equation}
\mathcal{V}_s' = \mathcal{V}_m + \boldsymbol{d}.
\end{equation}
This space encompasses both the pre-existing spontaneous Hall response (encoded in $\mathcal{V}_m$) and the component parallel to the applied field. Since a magnetic field along $\boldsymbol{d}$ generally induces a net magnetization in this direction, we have $\boldsymbol{d} \in \mathcal{V}_m'$, and thus $\mathcal{V}_s' \subseteq \mathcal{V}_m'$. Field-induced IPAHE is characterized by a Hall response lying outside $\mathcal{V}_s'$, requiring a direction orthogonal to both the initial response and the field. The strict dimensional mismatch,
\begin{equation}
\dim(\mathcal{V}_m') > \dim(\mathcal{V}_s'),
\label{eq:field_ipah1}
\end{equation}
constitutes the symmetry condition for field-induced IPAHE. Here, $\dim(\mathcal{V}_s') = \dim(\mathcal{V}_m) + \alpha$, where $\alpha = 1$ if $\boldsymbol{d} \notin \mathcal{V}_m$ and $\alpha = 0$ otherwise.


Before applying Eq.~\ref{eq:field_ipah1},
we first examine how the symmetry operations within the 122 MPGs constrain an axial magnetization vector $\boldsymbol{M}$. These operations can be divided into three classes based on their effect on the dimensionality of the allowed magnetization space, $\dim(\mathcal{V}_m)$.
First, time-reversal symmetry $\mathcal{T}$ acts as $-\mathbb{I}$ in spin space and completely forbids any net magnetization, yielding $\dim(\mathcal{V}_m)=0$.  Second, pure spatial operations---namely $n$-fold rotations $C_{n\boldsymbol{d}}$ ($n\ge 2$) about an axis $\boldsymbol{d}$, rotation-inversions $\bar{C}_{n\boldsymbol{d}}$, and mirror planes $m_{\boldsymbol{d}}$ with normal $\boldsymbol{d}$---force the magnetization to be collinear with the rotation axis or mirror normal. These operations enforce zero net magnetization along the two directions perpendicular to $\boldsymbol{d}$, giving $\dim(\mathcal{V}_m)=1$. Third, antiunitary operations combining time reversal with a spatial operation---$C_{n\boldsymbol{d}}\mathcal{T}$ and $m_{\boldsymbol{d}}\mathcal{T}$---leave the two directions perpendicular to $\boldsymbol{d}$ invariant while forbidding the component along $\boldsymbol{d}$. This results in $\dim(\mathcal{V}_m)=2$. Inversion $I$, by contrast, does not constrain an axial vector and can be neglected in this analysis.

We now analyze the necessary conditions for Eq.~\eqref{eq:field_ipah1} to hold. Regardless of the initial state, the constrained polarization space always has dimension $\dim(\mathcal{V}_s') \ge 1$. For nonmagnetic or antiferromagnetic systems where $\dim(\mathcal{V}_m)=0$, the field direction is not contained in $\mathcal{V}_m$, so $\alpha=1$ and $\dim(\mathcal{V}_s')=1$. For ferromagnetic systems where $\dim(\mathcal{V}_m)\ge 1$, we have $\dim(\mathcal{V}_s') \ge \dim(\mathcal{V}_m) \ge 1$. Therefore, the condition $\dim(\mathcal{V}_m') > \dim(\mathcal{V}_s')$ necessarily requires $\dim(\mathcal{V}_m') \ge 2$.
This immediately implies that the field-lowered MPG $G_m'$ cannot contain any pure spatial rotation $C_{n\boldsymbol{d}}$ ($n\ge 2$) or mirror plane $m_{\boldsymbol{d}}$---the only pure spatial operations that leave the field invariant---because such operations would rigidly confine the magnetization to the field direction $\boldsymbol{d}$, forcing $\dim(\mathcal{V}_m')=1$. Consequently, $G_m'$ must belong to one of the MPGs that permit a two- or three-dimensional response space: $1$, $\bar{1}$, $2'$, $m'$, or $2'/m'$.

Therefore, IPAHE requires that the applied magnetic field must destroy all pure rotation and mirror symmetries of the initial MPG. We use $\boldsymbol{p}$ denotes a direction perpendicular to the field ($\boldsymbol{p} \perp \boldsymbol{d}$). A rotation $C_{n\boldsymbol{p}}$ or mirror plane $m_{\boldsymbol{p}}$, whose axis or normal is perpendicular to $\boldsymbol{d}$, is automatically broken by the field, whereas a rotation $C_{n\boldsymbol{d}}$ or mirror plane $m_{\boldsymbol{d}}$ about the field direction would be preserved; the above condition therefore amounts to requiring that $G_m$ possess no $C_{n\boldsymbol{d}}$ or $m_{\boldsymbol{d}}$ symmetry. Once the $C_{n\boldsymbol{p}}$ and $m_{\boldsymbol{p}}$ symmetries are broken, the two directions perpendicular to $\boldsymbol{p}$---previously forbidden by symmetry---become allowed for the net magnetization. The response space $\mathcal{V}_m'$ thus increases by two dimensions compared with $\mathcal{V}_m$. Meanwhile, the constrained space $\mathcal{V}_s'$ increases by one dimension (the field direction $\boldsymbol{d}$ itself). It is therefore natural to achieve the dimensional mismatch $\dim(\mathcal{V}_m') > \dim(\mathcal{V}_s')$. In this sense, field-induced IPAHE can be understood as a process in which the magnetic field destroys the rotation and mirror symmetries.

By contrast, the antiunitary operations behave differently. Those whose rotation axis or mirror normal is parallel to the field, $C_{n\boldsymbol{d}}\mathcal{T}$ and $m_{\boldsymbol{d}}\mathcal{T}$, constrain $\boldsymbol{M}$ to the plane perpendicular to $\boldsymbol{d}$ [$\dim(\mathcal{V}_m)=2$]; they are broken precisely by a field applied along $\boldsymbol{d}$, which releases the magnetization component along the field direction itself. Therefore, breaking the $C_{n\boldsymbol{d}}\mathcal{T}$ and $m_{\boldsymbol{d}}\mathcal{T}$ symmetries cannot induce a Hall response orthogonal to the applied magnetic field.

\begin{table*}
\centering
\setlength\tabcolsep{12pt}
\renewcommand{\arraystretch}{2.7}
\caption{Results of field-induced IPAHE. The first column lists the crystal system to which the MPGs belong. The second column lists the specific MPGs that permit field-induced IPAHE. The third column indicates the magnetic type, categorized as ferromagnet (no primes in the point group symbol) or non-ferromagnet (with primes or containing $.1'$). The fourth column gives the direction(s) of the applied magnetic field $\bm{B}$ ($x$, $y$, or $z$) under which IPAHE is allowed. The fifth column gives the total number of magnetic point groups listed in the corresponding row.}
	    \label{tab:IPAHE_results}
	    \begin{tabular}{|c|c|c|c|c|}
	        \hline
	        Crystal System & MPGs & Type & \makecell{Magnetic Field \\ Direction} & Count \\
	        \hline
	        Triclinic & $1.1', \bar{1}.1', \bar{1}'$ & Non-ferromagnet & $\bm{B}||x$, or $\bm{B}||y$, or $\bm{B}||z$ & 3 \\
	        \hline
	        \multirow{2}{*}{Monoclinic} & $2, m, 2/m$ & Ferromagnet & $\bm{B}||x$ or $\bm{B}||z$ & 3 \\
	        \cline{2-5}
	        & $2.1', m.1', 2/m.1', 2'/m, 2/m'$ & Non-ferromagnet & $\bm{B}||x$ or $\bm{B}||z$ & 5 \\
	        \hline
	        \multirow{2}{*}{Tetragonal} & $4, \bar{4}, 4/m$ & Ferromagnet & $\bm{B}||x$ or $\bm{B}||y$ & 3 \\
	        \cline{2-5}
	        & \makecell{$4.1', 4', \bar{4}.1', \bar{4}',$ \\ $4/m.1', 4'/m, 4/m', 4'/m',$ \\ $4'm'm, \bar{4}'2'm, 4'/mm'm$} & Non-ferromagnet & $\bm{B}||x$ or $\bm{B}||y$ & 11 \\
	        \hline
	        \multirow{5}{*}{Trigonal} & $3, \bar{3}$ & Ferromagnet & $\bm{B}||x$ or $\bm{B}||y$ & 2 \\
	        \cline{2-5}
	        & $3.1', \bar{3}.1', \bar{3}'$ & Non-ferromagnet & $\bm{B}||x$ or $\bm{B}||y$ & 3 \\
	        \cline{2-5}
	        & \makecell{$32, 32.1', 3m, 3m.1', \bar{3}m, $ \\ $\bar{3}m.1', \bar{3}'m, \bar{3}'m' $} & Non-ferromagnet & $\bm{B}||y$ & 5 \\
	        \cline{2-5}
	        & $32', 3m', \bar{3}m'$ & Ferromagnet & $\bm{B}||x$ & 3 \\
	        \hline
	        \multirow{4}{*}{Hexagonal} & $6, \bar{6}, 6/m$ & Ferromagnet & $\bm{B}||x$ or $\bm{B}||y$ & 3 \\
	        \cline{2-5}
	        & \makecell{$6.1', 6', \bar{6}.1', \bar{6}',$ \\ $6/m.1', 6'/m, 6/m', 6'/m'$} & Non-ferromagnet & $\bm{B}||x$ or $\bm{B}||y$ & 8 \\
	        \cline{2-5}
	        & $6'22', 6'mm', \bar{6}'m2', 6'/m'mm'$ & Non-ferromagnet & $\bm{B}||y$ & 4 \\
	        \cline{2-5}
	        & $\bar{6}'m'2$ & Non-ferromagnet & $\bm{B}||x$ & 1 \\
	        \hline
	    \end{tabular}
\label{tab2}
\end{table*}

\subsection{C4. Results of field-induced IPAHE}
For each initial MPG $G_m$ and a given magnetic field direction $\boldsymbol{d}$, we first determine the dimensionality of the constrained polarization space, $\dim(\mathcal{V}_s')$. We then identify the dimensionality of the field-lowered response space, $\dim(\mathcal{V}_m')$, based on the obtained MPG $G_m'$. When $\dim(\mathcal{V}_m') > \dim(\mathcal{V}_s')$ is satisfied, field-induced IPAHE is realized. Furthermore, for each entry, we list the symmetry-allowed nonzero components of the Hall conductivity tensor $\sigma_{ij}$ based on $\dim(\mathcal{V}_m')$. The field-induced components that emerge solely due to the symmetry reduction induced by $\boldsymbol{B}$ are highlighted in red. The complete results for magnetic fields applied along the $x$, $y$, and $z$ directions are summarized in Tables~\ref{tab:s1}, \ref{tab:s2}, and \ref{tab:s3}, respectively.

We further summarize the key information in Table~\ref{tab2}, which lists all 54 MPGs (14 ferromagnetic and 40 non-ferromagnetic) that can host field-induced IPAHE, together with the required magnetic field directions. Among these, the triclinic system contributes 3 non-ferromagnetic MPGs ($1$, $1.1'$, $1'$), which allow IPAHE under any field direction ($x$, $y$, or $z$) owing to their low symmetry. The monoclinic system contributes 8 MPGs in total: 3 ferromagnetic ($2$, $m$, $2/m$) and 5 non-ferromagnetic ($2.1'$, $m.1'$, $2/m.1'$, $2'/m$, $2/m'$), all requiring the field to be applied along $x$ or $z$. The tetragonal system contributes 14 MPGs: 3 ferromagnetic ($4$, $\bar{4}$, $4/m$) and 11 non-ferromagnetic, all requiring the field to be perpendicular to the principal $z$ axis (i.e., along $x$ or $y$). The trigonal system contributes 13 MPGs: 5 ferromagnetic ($3$, $\bar{3}, 32', 3m', \bar{3}m'$) and 8 non-ferromagnetic. The hexagonal system contributes the largest number, 16 MPGs: 3 ferromagnetic ($6$, $\bar{6}$, $6/m$) and 13 non-ferromagnetic. Notably, no MPG in the orthorhombic or cubic crystal systems supports field-induced IPAHE.

We emphasize that our classification is exhaustive for the fixed Cartesian
Hall measurement geometries considered here, namely $\sigma_{xy}$,
$\sigma_{xz}$, and $\sigma_{yz}$. Although Table~\ref{tab2} lists only magnetic
fields applied along the Cartesian axes, the symmetry analysis covers all 122
MPGs and completely determines whether field-induced IPAHE can occur in these
measurement geometries for an arbitrarily oriented magnetic field. Accordingly,
the remaining 68 MPGs, which do not support IPAHE for fields along $x$, $y$, or
$z$, cannot realize IPAHE for any field direction within the fixed Hall
measurement configurations considered here.

\begin{table}[t]
\caption{\label{tab:ops}
Generic plane-preserving operations $g=(\mathcal R_g,p_g)$ for a field
rotating in the plane perpendicular to $\hat a$ [Eq.~(\ref{eq:brot})],
together with the transformed field angle $\phi_g$ and the scalar factor
$\eta_a$ [Eq.~(\ref{eq:rule})].
Here, $C_{na}$ and $S_{na}=M_aC_{na}$ denote a rotation and a
rotoreflection about the plane normal $\hat a$, respectively;
$M_a$ is the mirror perpendicular to $\hat a$;
$C_2(\phi_0)$ is a twofold rotation about an in-plane axis at angle
$\phi_0$; and $m_v(\phi_0)$ is a mirror containing $\hat a$, whose
in-plane normal points along $\phi_0$. Operations that do not preserve
the field plane impose no constraint on the angular function.}
\begin{ruledtabular}
\begin{tabular}{lccc}
operation $g$ & $(\mathcal R_g)_{aa}$ & $\phi_g$ & $\eta_a$ \\ \hline
$C_{na}$, $n=2,3,4,6$
    & $+1$ & $\phi+2\pi/n$ & $+1$ \\
$C_{na}\T$
    & $+1$ & $\phi+2\pi/n+\pi$ & $-1$ \\
$S_{na}$, $n=4,6$
    & $-1$ & $\phi+2\pi/n+\pi$ & $+1$ \\
$S_{na}\T$
    & $-1$ & $\phi+2\pi/n$ & $-1$ \\
$M_a$
    & $-1$ & $\phi+\pi$ & $+1$ \\
$M_a\T$
    & $-1$ & $\phi$ & $-1$ \\
$I$
    & $-1$ & $\phi$ & $+1$ \\
$\T$
    & $+1$ & $\phi+\pi$ & $-1$ \\
$I\T$
    & $-1$ & $\phi+\pi$ & $-1$ \\
$C_2(\phi_0)$
    & $-1$ & $2\phi_0-\phi$ & $-1$ \\
$C_2(\phi_0)\T$
    & $-1$ & $2\phi_0-\phi+\pi$ & $+1$ \\
$m_v(\phi_0)$
    & $+1$ & $2\phi_0-\phi$ & $-1$ \\
$m_v(\phi_0)\T$
    & $+1$ & $2\phi_0-\phi+\pi$ & $+1$ \\
\end{tabular}
\end{ruledtabular}
\end{table}

\twocolumngrid
\onecolumngrid
\setlength{\LTcapwidth}{\textwidth}
\small
\setlength\tabcolsep{2pt}
\renewcommand{\arraystretch}{2.2}
\begin{longtable}{|c|c|c|}
\caption{Leading field-induced angular functions $\sigma_{xz}(\phi)$ for the
34 MPGs that support an $xz$-plane-field-induced Hall response. The angle
$\phi$ is measured from the $+x$ axis toward the $+z$ axis within the $xz$
plane, so that $\phi=0$ for $\mathbf B\parallel+x$ and $\phi=\pi/2$ for
$\mathbf B\parallel+z$.}\label{tab:xz}\\
\hline
\textbf{Crystal system} & \textbf{MPGs} & $\sigma_{xz}(\phi)$ \\
\hline
\endfirsthead
\multicolumn{3}{c}{\tablename\ \thetable{} -- continued from previous page}\\
\hline
\textbf{Crystal system} & \textbf{MPGs} & $\sigma_{xz}(\phi)$ \\
\hline
\endhead
\hline
\multicolumn{3}{r}{Continued on next page}\\
\endfoot
\hline
\endlastfoot

Triclinic
&
\makecell[c]{
$1.1'$, $\bar{1}.1'$, $\bar{1}'$
}
&
$a_{1}\cos\phi+b_{1}\sin\phi$ \\
\hline

Trigonal
&
\makecell[c]{
$3$, $3.1'$, $\bar{3}$, $\bar{3}.1'$, $\bar{3}'$
}
&
$a_{1}\cos\phi+b_{1}\sin\phi$ \\
\hline

Tetragonal
&
\makecell[c]{
$4$, $4.1'$, $4'$, $\bar{4}$, $\bar{4}.1'$, $\bar{4}'$,\\
$4/m$, $4/m.1'$, $4'/m$, $4/m'$, $4'/m'$
}
&
$a_{1}\cos\phi$ \\
\hline

Hexagonal
&
\makecell[c]{
$6$, $6.1'$, $6'$, $\bar{6}$, $\bar{6}.1'$, $\bar{6}'$,\\
$6/m$, $6/m.1'$, $6'/m$, $6/m'$, $6'/m'$
}
&
$a_{1}\cos\phi$ \\
\hline

Trigonal
&
\makecell[c]{
$32'$, $3m'$, $\bar{3}m'$
}
&
$b_{1}\sin\phi$ \\
\hline

Hexagonal
&
$\bar{6}'m'2$
&
$a_{2}\cos2\phi$ \\
\hline

\end{longtable}

\section*{D. Field-direction dependence of the Hall conductance}

In this section, we derive the angular dependence of the Hall conductance
when a magnetic field of fixed magnitude is rotated within a given plane.

\subsection*{D1. Symmetry constraints on the Hall conductance}

A magnetic-point-group operation is denoted by
$g=(\mathcal{R}_g,p_g)$, where $\mathcal{R}_g$ is its spatial part and
$p_g=0$ ($p_g=1$) denotes a unitary (antiunitary) operation. Both the
magnetic field $\bm B$ and the Hall vector
$\boldsymbol{\sigma}^{\mathrm{H}}
=
\left(
\sigma_{yz},
\sigma_{zx},
\sigma_{xy}
\right)$ are time-reversal-odd axial vectors. Their transformation laws are
therefore
\begin{equation}
\bm B
\longrightarrow
\mathcal{M}_g\bm B,
\qquad
\boldsymbol{\sigma}^{\mathrm{H}}
\longrightarrow
\mathcal{M}_g\boldsymbol{\sigma}^{\mathrm{H}},
\label{eq:mlaw}
\end{equation}
where $\mathcal{M}_g
=
(-1)^{p_g}
\det(\mathcal{R}_g)\mathcal{R}_g$.

Let the magnetic field rotate in the plane spanned by two orthonormal
vectors $\hat{\bm u}$ and $\hat{\bm v}$:
\begin{equation}
\bm B(\phi)
=
B\left(
\cos\phi\,\hat{\bm u}
+
\sin\phi\,\hat{\bm v}
\right),
\qquad
\hat{\bm a}
=
\hat{\bm u}\times\hat{\bm v},
\label{eq:brot}
\end{equation}
where $B\geq0$ and $\phi$ is measured from $\hat{\bm u}$. The Hall
component normal to the rotation plane is
\begin{equation}
\sigma_H
=
\hat{\bm a}\cdot\boldsymbol{\sigma}^{\mathrm H}.
\label{eq:sigmaHdef}
\end{equation}
For the ordered planes $xy$, $yz$, and $zx$, this definition gives
\begin{equation}
\sigma_H=\sigma_{xy},
\qquad
\sigma_H=\sigma_{yz},
\qquad
\sigma_H=\sigma_{zx}=-\sigma_{xz},
\end{equation}
respectively.

A symmetry operation relates two fields within the same rotation plane only
if it preserves that plane. This requires
\begin{equation}
\mathcal{R}_g\hat{\bm a}
=
s_a\hat{\bm a},
\qquad
s_a=\pm1,
\label{eq:planepreserve}
\end{equation}
or equivalently,
\begin{equation}
(\mathcal{R}_g)_{aa}=s_a=\pm1.
\end{equation}
Since $\mathcal{R}_g$ is orthogonal, its in-plane and normal components then
decouple. Operations that do not satisfy
Eq.~\eqref{eq:planepreserve} map a generic in-plane field out of the
rotation plane and therefore impose no constraint on the angular scan
considered here.

For a plane-preserving operation, define $\phi_g$ by
\begin{equation}
\mathcal{M}_g\bm B(\phi)
=
\bm B(\phi_g).
\label{eq:phigdef}
\end{equation}
 The normal Hall component obeys
\begin{equation}
\sigma_H(\phi_g)
=
\eta_a(g)\sigma_H(\phi),
\label{eq:rule}
\end{equation}
with
\begin{equation}
\eta_a(g)
=
(-1)^{p_g}
\det(\mathcal{R}_g)
(\mathcal{R}_g)_{aa}
=
\pm1.
\label{eq:etaa}
\end{equation}
Thus, $\phi_g$ is determined by the in-plane action of
$\mathcal{M}_g$, whereas the sign of the normal Hall response is determined
by $\eta_a(g)$. Table~\ref{tab:ops} lists the corresponding
$\phi_g$ and $\eta_a(g)$ for the relevant plane-preserving operations.

Generally, the Hall
conductance $\sigma_H(\phi)$ under a fixed magnitude of magnetic field can generally be expanded as
\begin{align}
\sigma_H(\phi)
&=\sum_{n=1}^{\infty}\bigl[a_n\cos(n\phi)+b_n\sin(n\phi)\bigr].
\label{eq:fourier}
\end{align}
where we have fixed the magnitude of $B$. The leading harmonic of $\sigma_H(\phi)$ is completely
determined by the symmetry constraints imposed by Eq.~\ref{eq:rule}.

\twocolumngrid
\onecolumngrid
\setlength{\LTcapwidth}{\textwidth}
\small
\setlength\tabcolsep{2pt}
\renewcommand{\arraystretch}{2.4}
\begin{longtable}{|c|c|c|}
\caption{Leading field-induced angular functions $\sigma_{yz}(\phi)$ for the
45 MPGs that support a $yz$-plane-field-induced Hall response. The angle
$\phi$ is measured from the $+y$ axis toward the $+z$ axis within the $yz$
plane, so that $\phi=0$ for $\mathbf B\parallel+y$ and $\phi=\pi/2$ for
$\mathbf B\parallel+z$. }\label{tab:yz}\\
\hline
\textbf{Crystal system} & \textbf{MPGs} & $\sigma_{yz}(\phi)$ \\
\hline
\endfirsthead
\multicolumn{3}{c}{\tablename\ \thetable{} -- continued from previous page}\\
\hline
\textbf{Crystal system} & \textbf{MPGs} & $\sigma_{yz}(\phi)$ \\
\hline
\endhead
\hline
\multicolumn{3}{r}{Continued on next page}\\
\endfoot
\hline
\endlastfoot

Triclinic
&
\makecell[c]{
$1.1'$, $\bar{1}.1'$, $\bar{1}'$
}
&
$a_{1}\cos\phi+b_{1}\sin\phi$ \\
\hline

Monoclinic
&
\makecell[c]{
$2$, $2.1'$, $m$, $m.1'$,\\
$2/m$, $2/m.1'$, $2'/m$, $2/m'$
}
&
$b_{1}\sin\phi$ \\
\hline

Tetragonal
&
\makecell[c]{
$4$, $4.1'$, $4'$, $\bar{4}$, $\bar{4}.1'$, $\bar{4}'$,\\
$4/m$, $4/m.1'$, $4'/m$, $4/m'$, $4'/m'$
}
&
$a_{1}\cos\phi$ \\
\hline

Trigonal
&
\makecell[c]{
$3$, $3.1'$, $\bar{3}$, $\bar{3}.1'$, $\bar{3}'$
}
&
$a_{1}\cos\phi+b_{1}\sin\phi$ \\
\hline

Trigonal
&
\makecell[c]{
$32$, $3m$, $\bar{3}m$
}
&
$a_{2}\cos2\phi+b_{2}\sin2\phi$ \\
\hline

Hexagonal
&
\makecell[c]{
$6$, $6.1'$, $6'$, $\bar{6}$, $\bar{6}.1'$, $\bar{6}'$,\\
$6/m$, $6/m.1'$, $6'/m$, $6/m'$, $6'/m'$
}
&
$a_{1}\cos\phi$ \\
\hline

Hexagonal
&
\makecell[c]{
$6'22'$, $6'mm'$, $\bar{6}'m2'$, $6'/m'mm'$
}
&
$a_{2}\cos2\phi$ \\
\hline

\end{longtable}

\section*{D2. Results}
In the 54 MPGs that support IPAHE, we find that 28 MPGs, 34 MPGs, and 45 can support nonzero $\sigma_{xy}$, $\sigma_{xz}$, and  $\sigma_{yz}$ for the applied magnetic field in the $xy$-, $xz$-, and $yz$-plane. In the main text, we have presented the leading angular dependences of the field-induced Hall conductance $\sigma_{xy}(\phi)$. The results for $\sigma_{xz}(\phi)$, $\sigma_{yz}(\phi)$ is summarized in Tables.~\ref{tab:xz} and \ref{tab:yz}, respectively.

\section{E. Candidate materials for IPAHE}
\label{app:candidate_materials}

In this section, we present candidate materials for both spontaneous and field-induced IPAHE.

\subsection{E1. Candidate materials for spontaneous IPAHE}
Spontaneous IPAHE in ferromagnets requires the dimensional mismatch $\dim(\mathcal{V}_m) > \dim(\mathcal{V}_s) > 0$. Since $\dim(\mathcal{V}_s) \ge 1$ for any ferromagnet, this necessitates $\dim(\mathcal{V}_m) \ge 2$. Among the 31 ferromagnetic MPGs, only five satisfy this requirement: $2'$, $m'$, and $2'/m'$ with $\dim(\mathcal{V}_m)=2$, and $1$ and $\bar{1}$ with $\dim(\mathcal{V}_m)=3$.
For these five groups, the mismatch condition imposes distinct constraints on $\dim(\mathcal{V}_s)$. In $2'$, $m'$, and $2'/m'$, where $\dim(\mathcal{V}_m)=2$, spontaneous IPAHE requires $\dim(\mathcal{V}_s)=1$. This corresponds to a net magnetization along a single direction, which is most naturally realized in collinear ferromagnets, though certain canted or noncoplanar configurations with a uniaxial net magnetization are also permitted. In $1$ and $\bar{1}$, where $\dim(\mathcal{V}_m)=3$, the condition requires $\dim(\mathcal{V}_s)=1$ or $2$, allowing for collinear, coplanar, or noncoplanar magnetic orders provided that the net magnetization does not span all three dimensions. To identify concrete material realizations, we have systematically screened the ferromagnetic compounds cataloged in the MAGNDATA database \cite{GallegoMagndataI}. Table~\ref{tabs5} presents the 33 candidate materials identified from this screening.

\subsection{E2. Candidate materials for field-induced IPAHE}

In Table~\ref{tab2}, we list all 54 MPGs capable of supporting field-induced IPAHE. Based on these target MPGs, candidate materials can be identified from magnetic structure databases. We specifically focus on magnetic materials, thereby excluding nonmagnetic compounds from our screening.

We first consider materials characterized by ferromagnetic point groups, which comprise 14 MPGs: $2$, $m$, and $2/m$ in the monoclinic system; $4$, $\bar{4}$, and $4/m$ in the tetragonal system; $3$, $\bar{3}$, $32'$, $3m'$, and $\bar{3}m'$ in the trigonal system; and $6$, $\bar{6}$, and $6/m$ in the hexagonal system. In these ferromagnetic materials, a spontaneous magnetization along a single direction already exists at zero field. Applying an external magnetic field along the symmetry-allowed directions (specified in Table~\ref{tab2}) can then induce the IPAHE.
These MPGs can describe both conventional ferromagnets and spin-orbital magnets. In Tables~\ref{tabs8} and \ref{tabs6}, we list the candidate materials for realizing field-induced IPAHE in these two classes, respectively. While spin-orbital magnets exhibit antiferromagnetic order, they generally display nonrelativistic spin splitting due to the breaking of $P\mathcal{T}$ symmetry. These materials belong to the category of unconventional magnetism, including altermagnets with collinear magnetic order \cite{ifmmode2022}. Moreover, unconventional magnetism has been completely classified into four classes according to their parity properties, namely even-parity, odd-parity, hybrid-parity, and unconstrained parity \cite{2026arXiv260307643L,2026arXiv260521336L}. These novel magnetic phases, in addition to altermagnets, can also support IPAHE.

For the remaining 40 non-ferromagnetic MPGs, 16 belong to gray groups (Type-II), which can describe either nonmagnetic materials or magnetic materials with $\mathcal{T}\tau$ symmetry. Among these, the three MPGs $2/m.1'$, $4/m.1'$, and $6/m.1'$ preserve $P\mathcal{T}$ symmetry, enforcing spin degeneracy in the energy bands. This degeneracy is broken by the applied magnetic field, which can generate a Hall response orthogonal to the field direction. Here, we focus on magnetic materials with $\mathcal{T}\tau$ symmetry. In Table~\ref{tabs10}, we present candidate materials belonging to the target MPGs that can support IPAHE. We emphasize that in coplanar or noncoplanar magnetic orders where inversion symmetry is broken, the preserved $\mathcal{T}\tau$ symmetry enforces time-reversal-preserving non-relativistic spin splitting, placing these materials in the category of odd-parity magnets \cite{2026arXiv260307643L}.

In Table~\ref{tabs10}, we also present candidate antiferromagnetic materials belonging to the remaining 24 Type-III MPGs. Among these, $2'/m$, $2/m'$, $4/m'$, $6'/m$, and $6/m'$ preserve $P\mathcal{T}$ symmetry, which enforces spin degeneracy. For the remaining 19 MPGs, $P\mathcal{T}$ symmetry is broken, and these systems can generally exhibit nonrelativistic spin splitting. Specifically, in collinear magnetic orders, nonrelativistic spin splitting arises when the spin-contrast sublattices are related by rotation or mirror symmetry rather than by inversion or time reversal, realizing altermagnetism. In coplanar or noncoplanar magnetic orders, nonrelativistic spin splitting exists as long as there is no multi-spin rotation-translation symmetry, realizing unconventional magnetism. In Table~\ref{tabs10}, we present information on whether the candidate materials are unconventional magnets, including altermagnets, odd-parity magnets, and hybrid-parity magnets \cite{2026arXiv260307643L,2026arXiv260521336L}.

\section{F. Details of first-principles calculations}

The magnetic structures of Ca$_2$NiOsO$_6$ and RuO$_2$ used in this work were
obtained from the MAGNDATA collection hosted by the Bilbao Crystallographic
Server \cite{GallegoMagndataI,PerezMatoBCS2015}. In the absence of an external
magnetic field, the directions and magnitudes of the magnetic moments were
assigned according to the corresponding magnetic CIF files provided by
MAGNDATA.

Ca$_2$NiOsO$_6$ crystallizes in the monoclinic space group $P2_1/n$
(No.~14), and its magnetic structure has the MPG $2'/m'$.
In the noncollinear spin-orbit calculations, the magnetic moments were
constrained along the crystallographic $x$ direction, with the Ni and Os
moments aligned antiparallel, as specified in the magnetic CIF file.

RuO$_2$ adopts the tetragonal rutile structure with space group $P4_2/mnm$
(No.~136), and its magnetic structure has the MPG
$4'/mm'm$. In this structure, the Ru moments are antiferromagnetic
order along the $z$ direction. To simulate an external magnetic field
applied along the $x$ direction, a small uniform $x$ component of
$0.1~\mu_B$ was added to each Ru moment. This setup mimics the field-induced
canting of the antiferromagnetic moments while preserving the staggered
$z$-axis antiferromagnetic component.

First-principles calculations based on density functional theory were
performed using the Vienna \textit{Ab initio} Simulation Package
(VASP) \cite{KresseFurthmuller1996,KresseJoubert1996}. Core--valence
interactions were described using the projector augmented-wave (PAW)
method \cite{KresseG1999}. The exchange--correlation functional was treated
within the generalized gradient approximation (GGA) using the
Perdew--Burke--Ernzerhof (PBE) parametrization
\cite{PerdewBurkeErnzerhof1996}. Spin-orbit coupling (SOC) was included in
the noncollinear formalism, and crystallographic symmetry was switched off
during the self-consistent calculations in order to preserve the magnetic
configurations specified by the MAGNDATA magnetic CIF files. A kinetic-energy
cutoff of 550~eV was used for the plane-wave basis. The Brillouin zone (BZ)
was sampled with Monkhorst--Pack $\mathbf{k}$-point meshes corresponding to a
reciprocal-space spacing of approximately $2\pi\times0.01$~\AA$^{-1}$. The
electronic self-consistency criterion was set to $10^{-8}$~eV, and a Gaussian
smearing of 0.05~eV was employed.

On-site correlation effects were included using the rotationally invariant
GGA+$U$ method of Dudarev \textit{et al.} \cite{Dudarev1998}. For RuO$_2$, an
effective interaction parameter of $U_{\rm eff}=2.0$~eV was applied to the
Ru-$4d$ states, consistent with previous first-principles calculations for
antiferromagnetic RuO$_2$ \cite{LiangRuO2U2023}. For Ca$_2$NiOsO$_6$,
$U_{\rm eff}=6.2$~eV was applied to the Ni-$3d$ states, following earlier
LSDA/GGA+$U$ benchmark calculations for Ni-$3d$ oxides \cite{CaiNiOU2008}.
Previous first-principles studies of Ca$_2$NiOsO$_6$ were also used to guide
the correlated magnetic-calculation setup \cite{MorrowCa2NiOsO6}.

Tight-binding Hamiltonians were constructed from maximally localized Wannier
functions using \textsc{wannier90}
\cite{MarzariVanderbilt1997,Mostofi2014,MarzariReview2012}. For RuO$_2$, the
spinor Wannier basis was constructed from Ru-$d$ and O-$p$ orbitals. For
Ca$_2$NiOsO$_6$, a Ni-$d$/Os-$d$/O-$p$ basis was used in the final
magnetic-symmetry analysis, since both Ni and Os carry magnetic moments in
MAGNDATA entry 0.796.

The Berry curvature and anomalous Hall conductivity were evaluated using
\textsc{WannierTools} \cite{WuWannierTools2018} on the basis of the Wannier
tight-binding Hamiltonians. To restore the target magnetic symmetry in the
Wannier tight-binding model, the real-space Hamiltonian was symmetrized
using \texttt{wannhr\_symm\_Mag} \cite{YueWannhrSymmMag}.

\section{G. Effective two-band model for field-induced IPAHE}
\label{app:model}

To explicitly demonstrate the field-induced IPAHE in systems with a Berry curvature quadrupole, we construct a 2D two-band model,
\begin{equation}
\begin{split}
\mathcal{H}(\boldsymbol{k}) = {}& tk^2 + v(k_x\sigma_y - k_y\sigma_x) + m(k_x^2 - k_y^2)\sigma_z \\
& + B(\cos\phi\,\sigma_x+\sin\phi\,\sigma_y),
\end{split}
\label{eq:H0}
\end{equation}
where $\sigma_i$ ($i=x,y,z$) are the Pauli matrices acting on the spin space, $t$ is the effective mass parameter, $v$ characterizes the spin-orbit coupling, $m$ denotes the N\'eel-order-induced $d_{x^2-y^2}$ mass term, and $B$ is the Zeeman energy of the in-plane field $\boldsymbol{B}=B(\cos\phi,\sin\phi,0)$ with $\phi$ the azimuthal angle measured from the $x$ axis. At $h=0$, $\mathcal{H}$ belongs to the MPG $4'm'm$ and the Berry-curvature quadrupole for this model has been studied in Ref.~\cite{PhysRevB.107.115142}. In this case, the Hall conductivity vanishes enforced by the $C_{4z}\mathcal T$ symmetry.

Writing the Hamiltonian as $\mathcal H=\varepsilon_0\sigma_0+\boldsymbol d\cdot\boldsymbol\sigma$ with $\sigma_0$ the identity matrix, we have $\varepsilon_0(\boldsymbol k)=tk^2$ and
\begin{equation}
\boldsymbol d(\boldsymbol k)=\bigl[-vk_y+B\cos\phi,\;vk_x+B\sin\phi,\;m(k_x^2-k_y^2)\bigr],
\label{eq:d}
\end{equation}
so that the eigenenergies are $E_\pm(\boldsymbol k)=\varepsilon_0\pm|\boldsymbol d|$. The Berry curvature of band $n=\pm$ is $\Omega_{\pm,z}=\pm\,\boldsymbol d\cdot(\partial_{k_x}\boldsymbol d\times\partial_{k_y}\boldsymbol d)/(2|\boldsymbol d|^3)$, with the exact triple product
\begin{equation}
\boldsymbol d\cdot(\partial_{k_x}\boldsymbol d\times\partial_{k_y}\boldsymbol d)
= -mv^2(k_x^2-k_y^2)-2mvB\,(k_x\sin\phi+k_y\cos\phi),
\label{eq:triple}
\end{equation}
and the intrinsic Hall conductivity follows from
\begin{equation}
\sigma_{xy} = \frac{e^2}{\hbar}\int\frac{d^2k}{(2\pi)^2}\,f(E_-(\boldsymbol k))\,\Omega_{-,z}(\boldsymbol k),
\label{eq:sigmaxy}
\end{equation}
with $f$ the Fermi-Dirac distribution.

At $B=0$ the two bands touch at the $\Gamma$ point. The in-plane field does not gap this node linearly but shifts it to
\begin{equation}
\boldsymbol k_0=\frac{B}{v}\,(-\sin\phi,\;\cos\phi),
\label{eq:node}
\end{equation}
and generates a $\sigma_z$ mass at the shifted node,
\begin{equation}
M=m\bigl(k_{0x}^2-k_{0y}^2\bigr)=-\frac{mB^2}{v^2}\cos2\phi,
\label{eq:mass}
\end{equation}
whose sign is controlled by the field direction. The gapped Dirac cone at $\boldsymbol k_0$ carries the Berry flux $-\pi\,\mathrm{sgn}\,M$; since the total flux vanishes at $B=0$ by $C_{4z}\mathcal T$, the field-induced Hall conductivity for the chemical potential $\mu$  
located inside the energy gap is
\begin{equation}
\sigma_{xy}(\phi)= -\,\frac{e^2}{2h}\,\mathrm{sgn}\!\bigl(m\cos2\phi\bigr).
\label{eq:final}
\end{equation}
When the chemical potential $\mu$ is above the energy gap, the field-induced Hall conductivity is \cite{yrs7-m6zy}
\begin{equation}
\sigma_{xy}(\phi)= -\frac{e^2B^2m\cos(2\phi)}{2h\mu v^2}.
\label{eq:final}
\end{equation}
 Thus the in-plane field unlocks a finite Hall conductivity from the Berry-curvature quadrupole, whose angular dependence is governed by $\cos2\phi$.

\section{H. IPAHE under arbitrary-direction magnetic fields}
\label{app:arbitrary_field}

The classification in Table~\ref{tab2} assumes a conventional measurement
geometry in which the magnetic field is applied along one of the fixed
Cartesian axes and the Hall response is resolved into $\sigma_{xy}$,
$\sigma_{xz}$, and $\sigma_{yz}$. Although sufficient for this measurement
convention, such a classification does not exhaust all possible field
orientations and Hall-detection geometries.

As an example, we consider a magnetic field applied along the crystallographic
$[112]$ direction,
\begin{equation}
\bm B
=
B\hat{\bm d},
\qquad
\hat{\bm d}
=
\frac{\hat{\bm x}+\hat{\bm y}+2\hat{\bm z}}{\sqrt{6}}.
\label{eq:B112}
\end{equation}
Then the effective
spin-polarization space relevant to the fixed-axis measurement convention
is
\begin{equation}
\mathcal V_s^{'}
=
\operatorname{span}
\left\{
\hat{\bm x},
\hat{\bm y},
\hat{\bm z}
\right\},
\qquad
\dim\mathcal V_s^{'}=3.
\label{eq:Vs112Cartesian}
\end{equation}
Here, $\dim\mathcal V_s^{'}$ counts the number of nonzero Cartesian components of magnetic field in the fixed measurement basis, rather than the
intrinsic dimension of the line spanned by the single vector $\bm d$.
Consequently, none of the three Cartesian Hall-vector basis directions
associated with $\sigma_{yz}$, $\sigma_{zx}$, and $\sigma_{xy}$ is
strictly perpendicular to $\bm M$. Within this fixed-axis convention, these
three standard Hall channels therefore do not individually satisfy the
geometric condition for the IPAHE.

This restriction can be removed by choosing a measurement basis adapted to
the field direction. Let $\hat{\bm z}'=\hat{\bm d}$ and choose
$\hat{\bm x}'$ and $\hat{\bm y}'$ perpendicular to $\hat{\bm z}'$. In this
basis, the spin polarization has support only along
$\hat{\bm z}'$,
\begin{equation}
\widetilde{\mathcal V}_s^{'}
=
\operatorname{span}\{\hat{\bm z}'\},
\qquad
\dim\widetilde{\mathcal V}_s^{'}=1,
\label{eq:Vs112Rotated}
\end{equation}
whereas the Hall-vector components $\sigma_{y'z'}$ and $\sigma_{z'x'}$
are strictly perpendicular to $\bm M(\bm B)$. If either component is
symmetry allowed, the corresponding response satisfies the geometric
condition for field-induced IPAHE. Thus, allowing arbitrary field
orientations together with an adapted Hall-detection geometry may reveal
field-induced IPAHE in additional MPGs beyond those listed in
Table~\ref{tab2}.

\clearpage
\twocolumngrid
\onecolumngrid
\setlength{\LTcapwidth}{\textwidth}
\small
\setlength\tabcolsep{2pt}
\renewcommand{\arraystretch}{1}
\begin{longtable}{|c|c|c|c|c|c|c|c|c|c|}
\caption{Field-reduced symmetry and field-induced IPAHE for all 122 MPGs under the magnetic field along the $x$ direction. IPAHE is realized when $\dim(\mathcal{V}'_m) > \dim(\mathcal{V}'_s)$. "Crystal System" denotes the crystallographic system to which the point group belongs. "Initial MPG" is the MPG before applying the external magnetic field. "Type" indicates the type of MPG (Type-I: colorless, Type-II: gray, Type-III: black-and-white). "FM?" specifies whether the group is a ferromagnetic point group. "Surviving operations" lists the symmetry operations preserved under the applied magnetic field. "Final MPG" represents the reduced MPG after symmetry reduction by the field. "$\dim\mathcal{V}'_s$" is the dimensionality of the constrained polarization space. "$\dim\mathcal{V}'_m$" is the dimensionality of the response space. "$\sigma_{ij} \neq 0$" shows the symmetry-allowed nonzero components of the Hall conductivity tensor and the red components are associated with the field-induced IPAHE. "IPAHE?" indicates whether the IPAHE is supported (Yes) or not (No) under the specified field direction.}\label{tab:s1}\\
\hline
\textbf{Crystal System} & \textbf{Initial MPG} & \textbf{Type} & \textbf{FM?} & \textbf{Surviving operations} & \textbf{Final MPG} & $\mathbf{\dim\mathcal V_s'}$ & $\mathbf{\dim\mathcal V_m'}$ & $\sigma_{ij}\neq 0$ & \textbf{IPAHE?} \\
\hline
\endfirsthead
\multicolumn{10}{c}{\tablename\ \thetable{} -- continued from previous page}\\
\hline
\textbf{Crystal System} & \textbf{Initial MPG} & \textbf{Type} & \textbf{FM?} & \textbf{Surviving operations} & \textbf{Final MPG} & $\mathbf{\dim\mathcal V_s'}$ & $\mathbf{\dim\mathcal V_m'}$ & $\sigma_{ij}\neq 0$ & \textbf{IPAHE?} \\
\hline
\endhead
\hline
\multicolumn{10}{r}{Continued on next page}\\
\endfoot
\hline
\endlastfoot
 \multirow{5}{*}{Triclinic} & $1$ & I & Yes & $\{E\}$ & $1$ & 3 & 3 & $\sigma_{xy},\sigma_{xz},\sigma_{yz}$ & No \\
  & $1.1'$ & II & No & $\{E\}$ & $1$ & 1 & 3 & $\textcolor{red}{\sigma_{xy}},\textcolor{red}{\sigma_{xz}},\sigma_{yz}$ & Yes \\
  & $\bar{1}$ & I & Yes & $\{E, I\}$ & $\bar{1}$ & 3 & 3 & $\sigma_{xy},\sigma_{xz},\sigma_{yz}$ & No \\
  & $\bar{1}.1'$ & II & No & $\{E, I\}$ & $\bar{1}$ & 1 & 3 & $\textcolor{red}{\sigma_{xy}},\textcolor{red}{\sigma_{xz}},\sigma_{yz}$ & Yes \\
  & $\bar{1}'$ & III & No & $\{E\}$ & $1$ & 1 & 3 & $\textcolor{red}{\sigma_{xy}},\textcolor{red}{\sigma_{xz}},\sigma_{yz}$ & Yes \\
\hline
 \multirow{11}{*}{Monoclinic} & $2$ & I & Yes & $\{E\}$ & $1$ & 2 & 3 & $\textcolor{red}{\sigma_{xy}},\sigma_{xz},\sigma_{yz}$ & Yes \\
  & $2.1'$ & II & No & $\{E, C_{2y}'\}$ & $2'$ & 1 & 2 & $\textcolor{red}{\sigma_{xy}},\sigma_{yz}$ & Yes \\
  & $2'$ & III & Yes & $\{E, C_{2y}'\}$ & $2'$ & 2 & 2 & $\sigma_{xy},\sigma_{yz}$ & No \\
\hline
\end{longtable}

\newpage
\twocolumngrid
\onecolumngrid
\setlength{\LTcapwidth}{\textwidth}
\small
\setlength\tabcolsep{2pt}
\renewcommand{\arraystretch}{1}
\begin{longtable}{|c|c|c|c|c|c|c|c|c|c|}
\caption{Field-reduced symmetry and field-induced IPAHE for all 122 MPGs under the magnetic field along the $y$ direction. IPAHE is realized when $\dim(\mathcal{V}'_m) > \dim(\mathcal{V}'_s)$. "Crystal System" denotes the crystallographic system to which the point group belongs. "Initial MPG" is the MPG before applying the external magnetic field. "Type" indicates the type of MPG (Type-I: colorless, Type-II: gray, Type-III: black-and-white). "FM?" specifies whether the group is a ferromagnetic point group. "Surviving operations" lists the symmetry operations preserved under the applied magnetic field. "Final MPG" represents the reduced MPG after symmetry reduction by the field. "$\dim\mathcal{V}'_s$" is the dimensionality of the constrained polarization space. "$\dim\mathcal{V}'_m$" is the dimensionality of the response space. "$\sigma_{ij} \neq 0$" shows the symmetry-allowed nonzero components of the Hall conductivity tensor and the red components are associated with the field-induced IPAHE. "IPAHE?" indicates whether the IPAHE is supported (Yes) or not (No) under the specified field direction.}\label{tab:s2}\\
\hline
\textbf{Crystal System} & \textbf{Initial MPG} & \textbf{Type} & \textbf{FM?} & \textbf{Surviving operations} & \textbf{Final MPG} & $\mathbf{\dim\mathcal V_s'}$ & $\mathbf{\dim\mathcal V_m'}$ & $\sigma_{ij}\neq 0$ & \textbf{IPAHE?} \\
\hline
\endfirsthead
\multicolumn{10}{c}{\tablename\ \thetable{} -- continued from previous page}\\
\hline
\textbf{Crystal System} & \textbf{Initial MPG} & \textbf{Type} & \textbf{FM?} & \textbf{Surviving operations} & \textbf{Final MPG} & $\mathbf{\dim\mathcal V_s'}$ & $\mathbf{\dim\mathcal V_m'}$ & $\sigma_{ij}\neq 0$ & \textbf{IPAHE?} \\
\hline
\endhead
\hline
\multicolumn{10}{r}{Continued on next page}\\
\endfoot
\hline
\endlastfoot
 \multirow{5}{*}{Triclinic} & $1$ & I & Yes & $\{E\}$ & $1$ & 3 & 3 & $\sigma_{xy},\sigma_{xz},\sigma_{yz}$ & No \\
  & $1.1'$ & II & No & $\{E\}$ & $1$ & 1 & 3 & $\textcolor{red}{\sigma_{xy}},\sigma_{xz},\textcolor{red}{\sigma_{yz}}$ & Yes \\
  & $\bar{1}$ & I & Yes & $\{E, I\}$ & $\bar{1}$ & 3 & 3 & $\sigma_{xy},\sigma_{xz},\sigma_{yz}$ & No \\
  & $\bar{1}.1'$ & II & No & $\{E, I\}$ & $\bar{1}$ & 1 & 3 & $\textcolor{red}{\sigma_{xy}},\sigma_{xz},\textcolor{red}{\sigma_{yz}}$ & Yes \\
  & $\bar{1}'$ & III & No & $\{E\}$ & $1$ & 1 & 3 & $\textcolor{red}{\sigma_{xy}},\sigma_{xz},\textcolor{red}{\sigma_{yz}}$ & Yes \\
\hline
 \multirow{11}{*}{Monoclinic} & $2$ & I & Yes & $\{E, C_{2y}\}$ & $2$ & 1 & 1 & $\sigma_{xz}$ & No \\
  & $2.1'$ & II & No & $\{E, C_{2y}\}$ & $2$ & 1 & 1 & $\sigma_{xz}$ & No \\
  & $2'$ & III & Yes & $\{E\}$ & $1$ & 3 & 3 & $\sigma_{xy},\sigma_{xz},\sigma_{yz}$ & No \\
\hline
\end{longtable}

\clearpage
\twocolumngrid
\onecolumngrid
\setlength{\LTcapwidth}{\textwidth}
\small
\setlength\tabcolsep{2pt}
\renewcommand{\arraystretch}{1}
\begin{longtable}{|c|c|c|c|c|c|c|c|c|c|}
\caption{Field-reduced symmetry and field-induced IPAHE for all 122 MPGs under the magnetic field along the $z$ direction. IPAHE is realized when $\dim(\mathcal{V}'_m) > \dim(\mathcal{V}'_s)$. "Crystal System" denotes the crystallographic system to which the point group belongs. "Initial MPG" is the MPG before applying the external magnetic field. "Type" indicates the type of MPG (Type-I: colorless, Type-II: gray, Type-III: black-and-white). "FM?" specifies whether the group is a ferromagnetic point group. "Surviving operations" lists the symmetry operations preserved under the applied magnetic field. "Final MPG" represents the reduced MPG after symmetry reduction by the field. "$\dim\mathcal{V}'_s$" is the dimensionality of the constrained polarization space. "$\dim\mathcal{V}'_m$" is the dimensionality of the response space. "$\sigma_{ij} \neq 0$" shows the symmetry-allowed nonzero components of the Hall conductivity tensor and the red components are associated with the field-induced IPAHE. "IPAHE?" indicates whether the IPAHE is supported (Yes) or not (No) under the specified field direction.}\label{tab:s3}\\
\hline
\textbf{Crystal System} & \textbf{Initial MPG} & \textbf{Type} & \textbf{FM?} & \textbf{Surviving operations} & \textbf{Final MPG} & $\mathbf{\dim\mathcal V_s'}$ & $\mathbf{\dim\mathcal V_m'}$ & $\sigma_{ij}\neq 0$ & \textbf{IPAHE?} \\
\hline
\endfirsthead
\multicolumn{10}{c}{\tablename\ \thetable{} -- continued from previous page}\\
\hline
\textbf{Crystal System} & \textbf{Initial MPG} & \textbf{Type} & \textbf{FM?} & \textbf{Surviving operations} & \textbf{Final MPG} & $\mathbf{\dim\mathcal V_s'}$ & $\mathbf{\dim\mathcal V_m'}$ & $\sigma_{ij}\neq 0$ & \textbf{IPAHE?} \\
\hline
\endhead
\hline
\multicolumn{10}{r}{Continued on next page}\\
\endfoot
\hline
\endlastfoot
 \multirow{5}{*}{Triclinic} & $1$ & I & Yes & $\{E\}$ & $1$ & 3 & 3 & $\sigma_{xy},\sigma_{xz},\sigma_{yz}$ & No \\
  & $1.1'$ & II & No & $\{E\}$ & $1$ & 1 & 3 & $\sigma_{xy},\textcolor{red}{\sigma_{xz}},\textcolor{red}{\sigma_{yz}}$ & Yes \\
  & $\bar{1}$ & I & Yes & $\{E, I\}$ & $\bar{1}$ & 3 & 3 & $\sigma_{xy},\sigma_{xz},\sigma_{yz}$ & No \\
  & $\bar{1}.1'$ & II & No & $\{E, I\}$ & $\bar{1}$ & 1 & 3 & $\sigma_{xy},\textcolor{red}{\sigma_{xz}},\textcolor{red}{\sigma_{yz}}$ & Yes \\
  & $\bar{1}'$ & III & No & $\{E\}$ & $1$ & 1 & 3 & $\sigma_{xy},\textcolor{red}{\sigma_{xz}},\textcolor{red}{\sigma_{yz}}$ & Yes \\
\hline
 \multirow{11}{*}{Monoclinic} & $2$ & I & Yes & $\{E\}$ & $1$ & 2 & 3 & $\sigma_{xy},\sigma_{xz},\textcolor{red}{\sigma_{yz}}$ & Yes \\
  & $2.1'$ & II & No & $\{E, C_{2y}'\}$ & $2'$ & 1 & 2 & $\sigma_{xy},\textcolor{red}{\sigma_{yz}}$ & Yes \\
  & $2'$ & III & Yes & $\{E, C_{2y}'\}$ & $2'$ & 2 & 2 & $\sigma_{xy},\sigma_{yz}$ & No \\
\hline
\end{longtable}

	\clearpage
	\twocolumngrid
	\onecolumngrid
	\setlength{\LTcapwidth}{\textwidth}
	\small
	\setlength\tabcolsep{5pt}
	\renewcommand{\arraystretch}{2}
	\begin{longtable}{|c|c|c|c|c|c|c|c|c|}
	\caption{Candidate materials for spontaneous IPAHE.  The first column, ID, is the unique identifier for the candidate magnetic material. 
 The second column, Chemical formula, is the chemical composition of the material. The third column, OSSG, is the oriented spin space group of the materials. The fourth column, \textbf{MPG}, is the magnetic point group corresponding to the material. 
 The fifth column, $\boldsymbol{\mathcal{V}_s}$, gives the general vectors spanning the constrained polarization space. 
 The sixth column, $\boldsymbol{\mathcal{V}_m}$, gives the general vectors spanning the response space. 
The seventh column, $\boldsymbol{\dim\mathcal{V}_s}$, is the dimensionality of the constrained polarization space. 
 The eighth column, $\boldsymbol{\dim\mathcal{V}_m}$, is the dimensionality of the response space. 
The ninth column, $\boldsymbol{\sigma_{ij}\neq 0}$, lists the symmetry-allowed nonzero components of the anomalous Hall conductivity tensor (components in \textcolor{red}{red} are specifically highlighted associated with field-induced IPAHE).}\label{tabs5}\\
	\hline
	\textbf{ID} & \textbf{Chemical formula} & \textbf{OSSG} & \textbf{MPG} & $\boldsymbol{\mathcal{V}_s}$ & $\boldsymbol{\mathcal{V}_m}$ & $\boldsymbol{\dim\mathcal{V}_s}$ & $\boldsymbol{\dim\mathcal{V}_m}$ & $\boldsymbol{\sigma_{ij}\neq 0}$ \\
	\hline
	\endfirsthead
	\multicolumn{9}{c}{\tablename\ \thetable{} -- continued from previous page}\\
	\hline
	\textbf{ID} & \textbf{Chemical formula} & \textbf{OSSG} & \textbf{MPG} & $\boldsymbol{\mathcal{V}_s}$ & $\boldsymbol{\mathcal{V}_m}$ & $\boldsymbol{\dim\mathcal{V}_s}$ & $\boldsymbol{\dim\mathcal{V}_m}$ & $\boldsymbol{\sigma_{ij}\neq 0}$ \\
	\hline
	\endhead
	\hline
	\multicolumn{9}{r}{Continued on next page}\\
	\endfoot
	\hline
	\endlastfoot
	0.106 & DyVO$_3$ & $P^{m_{001}}n^{m_{010}}m^{2_{100}}a$ & $2'/m'$ & $(x,0,0)$ & $(x,0,z)$ & 1 & 2 & $\sigma_{yz},\textcolor{red}{\sigma_{xy}}$\\
	\hline
	0.121 & Li$_2$Co(SO$_4$)$_2$ & $P^{2_{100}}2_1/^{2_{100}}c^{m_{\alpha,0,\gamma}}1$ & $2'/m'$ & $(x,0,0)$ & $(x,0,z)$ & 1 & 2 & $\sigma_{yz},\textcolor{red}{\sigma_{xy}}$\\
	\hline
	0.139 & BiCrO$_3$ & $C^{1}2/^{1}c^{\infty_{0,\beta,\gamma}m}1$ & $\bar{1}$ & $(0,y,z)$ & $(x,y,z)$ & 2 & 3 & $\textcolor{red}{\sigma_{yz}},\sigma_{zx},\sigma_{xy}$\\
	\hline
	0.176 & Mn$_3$Ti$_2$Te$_6$ & $P^{1}\bar{3}^{1}c^{\infty_{100}m}1$ & $2'/m'$ & $(x,0,0)$ & $(x,0,z)$ & 1 & 2 & $\sigma_{yz},\textcolor{red}{\sigma_{xy}}$\\
	\hline
	0.233 & Mn$_2$FeSbO$_6$ & $R^{1}\bar{3}^{\infty_{100}m}1$ & $\bar{1}$ & $(x,0,0)$ & $(x,y,z)$ & 1 & 3 & $\sigma_{yz},\textcolor{red}{\sigma_{zx}},\textcolor{red}{\sigma_{xy}}$\\
	\hline
	0.234 & MnLaMnSbO$_6$ & $P^{1}4_2/^{1}n^{\infty_{100}m}1$ & $2'/m'$ & $(x,0,0)$ & $(x,y,0)$ & 1 & 2 & $\sigma_{yz},\textcolor{red}{\sigma_{zx}}$\\
	\hline
	0.258 & Li$_3$Fe$_2$(PO$_4$)$_3$ & $P^{1}2_1/^{1}c^{\infty_{100}m}1$ & $2'/m'$ & $(x,0,0)$ & $(x,y,0)$ & 1 & 2 & $\sigma_{yz},\textcolor{red}{\sigma_{zx}}$\\
	\hline
	0.297 & NaCrGe$_2$O$_6$ & $C^{1}2/^{1}c^{\infty_{001}m}1$ & $2'/m'$ & $(0,0,z)$ & $(x,0,z)$ & 1 & 2 & $\textcolor{red}{\sigma_{yz}},\sigma_{xy}$\\
\hline
	\end{longtable}

\clearpage
\twocolumngrid
\onecolumngrid
\setlength{\LTcapwidth}{\textwidth}
\small
\setlength\tabcolsep{5pt}
\renewcommand{\arraystretch}{1.6}
\begin{longtable}{|c|c|c|c|c|c|}
\caption{Candidate ferromagnets exhibiting field-induced IPAHE. The fourth column presents the direction of magnetic field that can give rise to IPAHE. The Hall conductance highlighted in red denotes the
component that is induced by the applied magnetic field.}\label{tabs8}\\
\hline
\textbf{ID} & \textbf{Formula} & \textbf{Zero-field MPG} & \textbf{$B$} & \textbf{Reduced MPG} & \textbf{$\sigma_{ij}\neq0$} \\
\hline
\endfirsthead
\multicolumn{6}{c}{\tablename\ \thetable{} -- continued from previous page}\\
\hline
\textbf{ID} & \textbf{Formula} & \textbf{Zero-field MPG} & \textbf{$B$} & \textbf{Reduced MPG} & \textbf{$\sigma_{ij}\neq0$} \\
\hline
\endhead
\hline \multicolumn{6}{r}{Continued on next page}\\
\endfoot
\hline
\endlastfoot
0.35 & Cu$_{2}$OSeO$_{3}$ & $3$ & \begin{tabular}[c]{@{}c@{}}$B\parallel x$\\$B\parallel y$\end{tabular} & \begin{tabular}[c]{@{}c@{}}$1$\\$1$\end{tabular} & \begin{tabular}[c]{@{}c@{}}$\sigma_{yz}, \textcolor{red}{\sigma_{zx}}, \textcolor{red}{\sigma_{xy}}$\\$\textcolor{red}{\sigma_{yz}}, \sigma_{zx}, \textcolor{red}{\sigma_{xy}}$\end{tabular} \\
\hline
0.39 & Nd$_{2}$NaRuO$_{6}$ & $2/m$ & \begin{tabular}[c]{@{}c@{}}$B\parallel x$\\$B\parallel z$\end{tabular} & \begin{tabular}[c]{@{}c@{}}$\bar{1}$\\$\bar{1}$\end{tabular} & \begin{tabular}[c]{@{}c@{}}$\sigma_{yz}, \textcolor{red}{\sigma_{zx}}, \textcolor{red}{\sigma_{xy}}$\\$\textcolor{red}{\sigma_{yz}}, \textcolor{red}{\sigma_{zx}}, \sigma_{xy}$\end{tabular} \\
\hline
0.60 & (NH$_{2}$(CH$_{3}$)$_{2}$)n(FeIIIFeII(HCOO)$_{6}$)n & $3m^{\prime}$ & $B\parallel x$ & $1$ & $\sigma_{yz}, \textcolor{red}{\sigma_{zx}}, \textcolor{red}{\sigma_{xy}}$ \\
\hline
0.64 & MnV$_{2}$O$_{4}$ & $4/m$ & \begin{tabular}[c]{@{}c@{}}$B\parallel x$\\$B\parallel y$\end{tabular} & \begin{tabular}[c]{@{}c@{}}$\bar{1}$\\$\bar{1}$\end{tabular} & \begin{tabular}[c]{@{}c@{}}$\sigma_{yz}, \textcolor{red}{\sigma_{zx}}, \textcolor{red}{\sigma_{xy}}$\\$\textcolor{red}{\sigma_{yz}}, \sigma_{zx}, \textcolor{red}{\sigma_{xy}}$\end{tabular} \\
\hline
0.77 & Tb$_{2}$Ti$_{2}$O$_{7}$ & $3m^{\prime}$ & $B\parallel x$ & $1$ & $\sigma_{yz}, \textcolor{red}{\sigma_{zx}}, \textcolor{red}{\sigma_{xy}}$ \\
\hline
0.78 & NiN$_{2}$O$_{6}$ & $3$ & \begin{tabular}[c]{@{}c@{}}$B\parallel x$\\$B\parallel y$\end{tabular} & \begin{tabular}[c]{@{}c@{}}$1$\\$1$\end{tabular} & \begin{tabular}[c]{@{}c@{}}$\sigma_{yz}, \textcolor{red}{\sigma_{zx}}, \textcolor{red}{\sigma_{xy}}$\\$\textcolor{red}{\sigma_{yz}}, \sigma_{zx}, \textcolor{red}{\sigma_{xy}}$\end{tabular} \\
\hline
\end{longtable}
\normalsize

\clearpage
\twocolumngrid
\onecolumngrid
\setlength{\LTcapwidth}{\textwidth}
\small
\setlength\tabcolsep{5pt}
\renewcommand{\arraystretch}{1.4}
\begin{longtable}{|c|c|c|c|c|c|c|}
\caption{Candidate spin orbital magnetism materials that can support field-induced IPAHE. The fourth column presents the direction of magnetic field that can give rise to IPAHE. The sixth column presents the category under the classification of unconventional magnets. The Hall conductance highlighted in red denotes the
component that is induced by the applied magnetic field.}\label{tabs6}\\
\hline
\textbf{ID} & \textbf{Formula} & \textbf{MPG} & \textbf{Field} & \textbf{Reduced MPG} & \textbf{phases} & $\boldsymbol{\sigma_{ij}\neq 0}$ \\
\hline
\endfirsthead
\multicolumn{7}{c}{\tablename\ \thetable{} -- continued from previous page}\\
\hline
\textbf{ID} & \textbf{Formula} & \textbf{MPG} & \textbf{Field} & \textbf{Reduced MPG} & \textbf{phases} & $\boldsymbol{\sigma_{ij}\neq 0}$ \\
\hline
\endhead
\hline
\multicolumn{7}{r}{Continued on next page}\\
\endfoot
\hline
\endlastfoot
0.61 & Li$_{2}$FeP$_{2}$O$_{7}$ & $2/m$ & $B\parallel x$ & $\bar1$ & even-parity magnet & $\sigma_{yz}$, $\sigma_{zx}$, \textcolor{red}{$\sigma_{xy}$}\\
       &  &  & $B\parallel z$ & $\bar1$ &  & \textcolor{red}{$\sigma_{yz}$}, $\sigma_{zx}$, $\sigma_{xy}$\\
\hline
0.83 & LiFeP$_{2}$O$_{7}$ & $2$ & $B\parallel x$ & $1$ & altermagnet & $\sigma_{yz}$, $\sigma_{zx}$, \textcolor{red}{$\sigma_{xy}$}\\
       &  &  & $B\parallel z$ & $1$ &  & \textcolor{red}{$\sigma_{yz}$}, $\sigma_{zx}$, $\sigma_{xy}$\\
\hline
0.99 & YBaMn$_{2}$O$_{5.5}$ & $2/m$ & $B\parallel x$ & $\bar1$ & altermagnet & $\sigma_{yz}$, $\sigma_{zx}$, \textcolor{red}{$\sigma_{xy}$}\\
       &  &  & $B\parallel z$ & $\bar1$ &  & \textcolor{red}{$\sigma_{yz}$}, $\sigma_{zx}$, $\sigma_{xy}$\\
\hline
0.105 & ErVO$_{3}$ & $2/m$ & $B\parallel x$ & $\bar1$ & altermagnet & $\sigma_{yz}$, $\sigma_{zx}$, \textcolor{red}{$\sigma_{xy}$}\\
       &  &  & $B\parallel z$ & $\bar1$ &  & \textcolor{red}{$\sigma_{yz}$}, $\sigma_{zx}$, $\sigma_{xy}$\\
\hline
\end{longtable}
\clearpage
\twocolumngrid

\clearpage
\twocolumngrid
\onecolumngrid
\setlength{\LTcapwidth}{\textwidth}
\small
\setlength\tabcolsep{5pt}
\renewcommand{\arraystretch}{1}
\begin{longtable}{|c|l|c|c|c|c|}
	\caption{Candidate materials belonging to non-ferromagnetic point groups that can support field-induced IPAHE. The fourth column presents the direction of the magnetic field that can give rise to IPAHE. The sixth column presents the category under the classification of unconventional magnets. The Hall conductance highlighted in red denotes the component that is induced by the applied magnetic field. AFM is the abbreviation of antiferromagnet.}\label{tabs10}\\
	\hline
	\textbf{ID} & \textbf{Formula} & \textbf{MPG} & $\boldsymbol{B}$ & $\boldsymbol{\sigma_{ij}\neq0}$ & \textbf{Magnetic class}\\
	\hline
	\endfirsthead
	\multicolumn{6}{c}{\tablename\ \thetable{} -- continued from previous page}\\
	\hline
	\textbf{ID} & \textbf{Formula} & \textbf{MPG} & $\boldsymbol{B}$ & $\boldsymbol{\sigma_{ij}\neq0}$ & \textbf{Magnetic class}\\
	\hline
	\endhead
	\hline
	\multicolumn{6}{r}{Continued on next page}\\
	\endfoot
	\hline
	\endlastfoot
	0.13 & $\mathrm{Ca}_{3}\mathrm{Co}_{2-x}\mathrm{Mn}_{x}\mathrm{O}_{6}$ & $3m$ & $B\parallel y$ & $\textcolor{red}{\sigma_{yz}}, \sigma_{zx}, \textcolor{red}{\sigma_{xy}}$ & altermagnet\\
	\hline
	0.15 & $\mathrm{Mn}\mathrm{F}_{2}$ & $4^{\prime}/mm^{\prime}m$ & $B\parallel x$ & $\sigma_{yz}, \textcolor{red}{\sigma_{xy}}$ & altermagnet\\
	0.15 & $\mathrm{Mn}\mathrm{F}_{2}$ & $4^{\prime}/mm^{\prime}m$ & $B\parallel y$ & $\sigma_{zx}, \textcolor{red}{\sigma_{xy}}$ & altermagnet\\
	\hline
	0.21 & $\mathrm{Pb}\mathrm{Ni}\mathrm{O}_{3}$ & $3m$ & $B\parallel y$ & $\textcolor{red}{\sigma_{yz}}, \sigma_{zx}, \textcolor{red}{\sigma_{xy}}$ & altermagnet\\
	\hline
	0.28 & $\mathrm{Li}\mathrm{Fe}\mathrm{Si}_{2}\mathrm{O}_{6}$ & $2/m^{\prime}$ & $B\parallel x$ & $\sigma_{yz}, \textcolor{red}{\sigma_{xy}}$ & AFM\\
	0.28 & $\mathrm{Li}\mathrm{Fe}\mathrm{Si}_{2}\mathrm{O}_{6}$ & $2/m^{\prime}$ & $B\parallel z$ & $\textcolor{red}{\sigma_{yz}}, \sigma_{xy}$ & AFM\\
	\hline
	0.29 & $\mathrm{Er}_{2}\mathrm{Ti}_{2}\mathrm{O}_{7}$ & $4^{\prime}/mm^{\prime}m$ & $B\parallel x$ & $\sigma_{yz}, \textcolor{red}{\sigma_{xy}}$ & AFM\\
	0.29 & $\mathrm{Er}_{2}\mathrm{Ti}_{2}\mathrm{O}_{7}$ & $4^{\prime}/mm^{\prime}m$ & $B\parallel y$ & $\sigma_{zx}, \textcolor{red}{\sigma_{xy}}$ & AFM\\
\hline
\end{longtable}
\clearpage
\twocolumngrid

\end{widetext}

\end{document}